\documentclass[preprintnumbers,s,nofootinbib,unsortedaddress,superscriptaddress,notitlepage,square,sort&compress,comma,numbers]{revtex4-1}
\usepackage{eurosym}
 \usepackage{lipsum}
\usepackage{amsfonts}
\usepackage{amsmath}
\usepackage{mathtools}
\usepackage{hyperref}
\usepackage{amssymb}
\usepackage[english]{babel}
\usepackage[utf8]{inputenc}
\usepackage{graphicx}
\usepackage{subcaption}
\usepackage{epsfig}
\usepackage{bm}
\usepackage{verbatim}
\usepackage{float}
\usepackage{wrapfig}
\usepackage{booktabs}
\usepackage{natbib}
\usepackage{multirow}
\usepackage[usenames, dvipsnames,x11names]{xcolor}
\usepackage{hyperref}
\usepackage{bm,bbm}
\usepackage{float}
\hypersetup{colorlinks=true, urlcolor=Turquoise4, linkcolor=Aquamarine, citecolor=Orchid4}
\numberwithin{equation}{section}

\newcommand{\mathsym}[1]{{}}

\begin{document}

\title{Littlest Inverse Seesaw Model.}
\author{A. E. C\'{a}rcamo Hern\'{a}ndez${}$}
\email{antonio.carcamo@usm.cl}
\author{S. F. King}
\email{king@soton.ac.uk}
\affiliation{$^{{a}}$Universidad T\'{e}cnica Federico Santa Mar\'{\i}a and Centro Cient%
\'{\i}fico-Tecnol\'{o}gico de Valpara\'{\i}so, \\
Casilla 110-V, Valpara\'{\i}so, Chile, $^{{b}}$School of Physics and
Astronomy, University of Southampton,\\
SO17 1BJ Southampton, United Kingdom }
\date{\today }

\begin{abstract}
We propose a minimal predictive inverse seesaw model based on two
right-handed neutrinos and two additional singlets, leading to the same low
energy neutrino mass matrix as in the Littlest Seesaw (LS) (type I) model.
In order to implement such a Littlest Inverse Seesaw (LIS) model, we have
used an $S_{4}$ family symmetry, together with other various symmetries,
flavons and driving fields. The resulting LIS model leads to an excellent
fit to the low energy neutrino parameters, including the prediction of a
normal neutrino mass ordering, exactly as in the usual LS model. However,
unlike the LS model, the LIS model allows charged lepton flavour violating
(CLFV) processes and lepton conversion in nuclei within reach of the forthcoming experiments.
\end{abstract}

\maketitle




\section{Introduction}

\label{Intro}

The existence of three fermion families, as well as their particular pattern
of masses and mixing angles is not explained in the Standard Model (SM), and
makes it appealing to consider a more fundamental theory addressing these
issues. This problem is especially challenging in the neutrino sector, where
the tiny values of the neutrino masses and large mixing angles between
generations suggest a different kind of underlying physics than what should
be responsible for the quark mass and mixing pattern. Whereas the small
quark mixing angles decrease from one generation to the next, in the lepton
sector two of the mixing angles are large, and one mixing angle is small.

The tiny neutrino masses might well originate from a type I seesaw mechanism~%
\cite%
{Minkowski:1977sc,Yanagida:1979as,GellMann:1980vs,Glashow:1979nm,Mohapatra:1979ia}%
, but in general this is hard to test experimentally. A minimal version of
the type I seesaw mechanism, involving just two right-handed neutrinos
(2RHN), was first proposed by one of us~\cite{King:1999mb,King:2002nf},
where we noted that the lightest neutrino is massless. Such a model with two
texture zeros in the Dirac neutrino mass matrix, proposed somewhat later~%
\cite{Frampton:2002qc}, is consistent with cosmological leptogenesis~\cite%
{Fukugita:1986hr,Guo:2003cc,Ibarra:2003up,Mei:2003gn,Guo:2006qa,Antusch:2011nz,Harigaya:2012bw,Zhang:2015tea}%
, but not compatible with the normal hierarchy (NH) of neutrino masses,
favoured by current data~\cite{Harigaya:2012bw, Zhang:2015tea}. On the other
hand the originally proposed 2RHN model with one texture zero~\cite%
{King:1999mb, King:2002nf}, actually predicts a NH.

The Littlest Seesaw (LS) model is a special case of 2RHN models with one
texture zero, which involves just two independent Yukawa couplings~\cite%
{King:2013iva,Bjorkeroth:2014vha,King:2015dvf,Bjorkeroth:2015ora,Bjorkeroth:2015tsa,King:2016yvg,Ballett:2016yod,King:2018fqh}%
, leading to a highly predictive scheme characterised by near maximal
atmospheric mixing and CP violation, with an approximate $\mu-\tau$
reflection symmetry~\cite{King:2018kka,King:2019tbt} but with additional
predictions arising from tri-maximal nature of the first column of the PMNS
matrix as well as a predicted reactor angle.

All type I seesaw models, including the LS model above, predict very tiny
branching ratios for the charged lepton flavor violating (LFV) decays, such
as $\mu\to e\gamma$, $\tau\to\mu\gamma$ and $\tau\to e\gamma$, several
orders of magnitude lower than their corresponding projective experimental
sensitivity. These very tiny branching ratios for the charged lepton flavor
violating (LFV) decays can be significantly enhanced by several orders of
magnitude if one considers low scale seesaw models \cite%
{Abada:2012cq,Deppisch:2004fa,Abada:2014kba,Abada:2014vea,Abada:2016awd,Abada:2018qok}%
. Thus if charged lepton flavour violating decays are observed in the
future, it will provide indubitable evidence of Physics Beyond the Standard
Model and their observation will shed light in the dynamics responsible for
the smallness of neutrino masses and the nature of lepton mixing.

In this paper, motivated by such considerations, we propose a fusion of the
LS model and the inverse seesaw model~\cite{Mohapatra:1986bd}, which we
refer to as the Littlest Inverse Seesaw (LIS) model. The neutrino mass
matrix of the LIS model, which involves two right-handed neutrinos plus two
additional singlets, is given by: 
\begin{equation}
M_{\nu }=\left( 
\begin{array}{ccc}
0_{3\times 3} & m_{D} & 0_{3\times 2} \\ 
m_{D}^{T} & 0_{2\times 2} & M \\ 
0_{2\times 3} & M^{T} & \mu 
\end{array}%
\right) ,  \label{Mnu0}
\end{equation}%
%
%
%
%
%
%
%
%
%
%
%
%
%
%
%
%
%
%
where $0_{n\times m}$ are $n\times m$ dimensional submatrices consisting of
all zeroes and the other submatrices in the flavour basis have the
structure: 
\begin{equation}
m_{D}\sim \left( 
\begin{array}{cc}
0 & b \\ 
a & 3b \\ 
a & b%
\end{array}%
\right) ,\hspace{0.7cm}\hspace{0.7cm}M\sim \left( 
\begin{array}{cc}
1 & 0 \\ 
0 & z%
\end{array}%
\right) ,\hspace{0.7cm}\hspace{0.7cm}\mu \sim \left( 
\begin{array}{cc}
1 & 0 \\ 
0 & \omega 
\end{array}%
\right) ,\hspace{0.7cm}\hspace{0.7cm}\omega =e^{\frac{2\pi i}{3}}.
\label{Mnublocks0}
\end{equation}%
The light active neutrino mass matrix arising from the inverse seesaw
formula $m_{\nu }=-m_{D}(M^{T})^{-1}\mu M^{-1}m_{D}^{T}$ takes the same form
as the usual LS model~\cite%
{King:2013iva,Bjorkeroth:2014vha,King:2015dvf,Bjorkeroth:2015ora,Bjorkeroth:2015tsa,King:2016yvg,Ballett:2016yod,King:2018fqh}%
: 
\begin{equation}
m_{\nu }=m_{\nu a}\left( 
\begin{array}{ccc}
0 & 0 & 0 \\ 
0 & 1 & 1 \\ 
0 & 1 & 1%
\end{array}%
\right) +m_{\nu b}\omega \left( 
\begin{array}{ccc}
1 & 3 & 1 \\ 
3 & 9 & 3 \\ 
1 & 3 & 1%
\end{array}%
\right)   \label{mnu0}
\end{equation}%
The above mass matrix structures are motivated by the phenomenological
success of the low energy mass matrix in Eq.~\ref{mnu0} which is identical
to that of the usual LS model, involving two right-handed neutrinos, but in
this case arising from the inverse seesaw model, including the two
additional singlets. Such an extension allows CLFV decays, such as $\mu
\rightarrow e\gamma $, at observable rates, since in the inverse seesaw
model small neutrino masses are explained by the smallness of the $\mu $
matrix \footnote{%
An example of a dynamical explanation for the smallness of the $\mu $
parameter of the inverse seesaw and its connection with Dark matter is
provided in Ref. \cite{Bertuzzo:2018ftf}}, which allows Dirac masses to be
large even for TeV scale values of $M$. This is the first low scale seesaw
model leading to a successful fit of the 6 physical observables of the
neutrino sector with only 2 effective free parameters. In our model the
small masses for the light active neutrinos are generated from an inverse
seesaw mechanism. In order to achieve the above mass matrices, we appeal to
standard approaches to the flavour puzzle based on symmetries, as follows.

The flavour puzzle of the SM indicates that New Physics has to be advocated
to explain the observed SM fermion mass and mixing pattern. This is the so
called flavour puzzle, which is not explained by the SM and provides
motivation for building models with additional scalars and fermions in their
particle spectrum and with extended symmetries which can be continuous or
discrete and their breaking produces the observed pattern of SM fermion mass
and mixing pattern. Several discrete groups have been employed in extensions
of the SM to tackle SM fermion flavor puzzle. In particular the discrete
group $S_{4}$ \cite{Altarelli:2009gn,Bazzocchi:2009da,Bazzocchi:2009pv,Toorop:2010yh,Patel:2010hr,Morisi:2011pm,Altarelli:2012bn,Mohapatra:2012tb,BhupalDev:2012nm,Varzielas:2012pa,Ding:2013hpa,Ishimori:2010fs,Ding:2013eca,Hagedorn:2011un,Campos:2014zaa,Dong:2010zu,VanVien:2015xha,deAnda:2017yeb,deAnda:2018oik,Chen:2019oey,deMedeirosVarzielas:2019cyj,deMedeirosVarzielas:2019hur,Chen:2019oey,CarcamoHernandez:2019kjy,King:2019vhv},
 together with the groups $A_{4}$ \cite%
{Ma:2001dn,He:2006dk,Feruglio:2008ht,Feruglio:2009hu,Chen:2009um,Varzielas:2010mp,Altarelli:2012bn,Ahn:2012tv,Memenga:2013vc,Felipe:2013vwa,Varzielas:2012ai, Ishimori:2012fg,King:2013hj,Hernandez:2013dta,Babu:2002dz,Altarelli:2005yx,Gupta:2011ct,Morisi:2013eca, Altarelli:2005yp,Kadosh:2010rm,Kadosh:2013nra,delAguila:2010vg,Campos:2014lla,Vien:2014pta,Joshipura:2015dsa,Hernandez:2015tna,Karmakar:2016cvb,Chattopadhyay:2017zvs,CarcamoHernandez:2017kra,Ma:2017moj,CentellesChulia:2017koy,Bjorkeroth:2017tsz,Srivastava:2017sno,Borah:2017dmk,Belyaev:2018vkl,CarcamoHernandez:2018aon,Srivastava:2018ser,delaVega:2018cnx,Borah:2018nvu,Pramanick:2019qpg,CarcamoHernandez:2019pmy,CarcamoHernandez:2019kjy}, $T_{7}$ \cite{Luhn:2007sy,Hagedorn:2008bc,Cao:2010mp,Luhn:2012bc,Kajiyama:2013lja,Bonilla:2014xla,Vien:2014gza, Vien:2015koa,Hernandez:2015cra,Arbelaez:2015toa}, $\Delta (27)$ \cite{Branco:1983tn,deMedeirosVarzielas:2006fc,Ma:2007wu,Varzielas:2012nn,Bhattacharyya:2012pi,Ferreira:2012ri,Ma:2013xqa,Nishi:2013jqa,Varzielas:2013sla,Aranda:2013gga,Harrison:2014jqa,Ma:2014eka,Abbas:2014ewa,Abbas:2015zna,Varzielas:2015aua,Bjorkeroth:2015uou,Chen:2015jta,Vien:2016tmh,Hernandez:2016eod,CarcamoHernandez:2017owh,deMedeirosVarzielas:2017sdv,Bernal:2017xat,CarcamoHernandez:2018iel,deMedeirosVarzielas:2018vab,CarcamoHernandez:2018hst} and $T^{\prime }$ \cite{Aranda:2000tm,Feruglio:2007uu,Sen:2007vx,Aranda:2007dp,Chen:2007afa,Eby:2008uc,Frampton:2008bz,Frampton:2008ep,Eby:2009ii,Frampton:2009fw,Eby:2011ph,Eby:2011qa,Chen:2011tj,Meroni:2012ty,Frampton:2013lva,Chen:2013wba,Girardi:2013sza,Carone:2016xsi,Vien:2018otl,Carone:2019lfc,CarcamoHernandez:2019vih,Ding:2019zxk}, is the smallest group containing an irreducible triplet representation
that can accommodate the three fermion families of the Standard model (SM).
These groups have been widely used in several extensions of the SM since
they are particular promising in providing a viable and predictive
description of the observed SM fermion mass spectrum and mixing parameters.
In the present article, we shall employ $S_4$, together with other auxiliary
symmetries, in order to achieve the above mass matrices of the LIS model,
together with a diagonal charged lepton mass matrix.

The current article is organized as follows. In section \ref{model} we
explain our model. In section \ref{leptonmassesandmixings} we present our
results in terms of neutrino masses and mixing. The implications of our
model in the lepton flavor violating decays $\mu\to e\gamma$, $%
\tau\to\mu\gamma$ and $\tau\to e\gamma$ and lepton conversion
in nuclei are studied in section \ref{leptonmassesandmixings}. We conclude in section \ref{conclusions}. A
description of the $S_4$ discrete group is presented in Appendix \ref{S4}.
The superpotential that determines the vacuum configuration for the $S_4$
doublet and triplet scalars of our model is presented in Appendix \ref{WS4}. 

\section{The model}

\label{model} We consider an $S_{4}$ flavour model for leptons where the
masses for the light active neutrinos are generated from an inverse seesaw
mechanism \cite{Mohapatra:1986bd,GonzalezGarcia:1988rw,Akhmedov:1995vm,Akhmedov:1995ip,Malinsky:2005bi,Malinsky:2009df,Abada:2014vea}. The implementation of the inverse seesaw mechanism in our model
relies in the inclusion of four gauge singlets right handed Majorana
neutrinos, which is the minimal amount of gauge singlet right handed
Majorana neutrinos needed to implement a realistic inverse seesaw mechanism
as pointed out for the first time in Ref. \cite{Malinsky:2009df}. The leptonic
and scalar spectrum of our model with their assigments under the $%
S_{4}\times U\left( 1\right) \times Z_{3}\times Z_{6}\times Z_{9}\times
U\left( 1\right) _{R}$ symmetry are shown in Table \ref{Themodel}.

The scalar spectrum of our model is composed of the $SU(2)_{L}$ Higgs doublets 
$H_{u}$, $H_{d}$ and several gauge singlet scalar fields, which are grouped
into one $S_{4}$ singlet, i.e., $\rho $, three $S_{4}$ doublets, i.e., $%
\varphi $, $\phi $, $\eta $ and five $S_{4}$ triplets, i.e., $\chi $, $\xi $%
, $\sigma _{\mu }$, $\sigma _{\tau }$, $\sigma _{e}$. The gauge singlet scalar fields $\sigma _{\mu }$, $\sigma _{\tau }$, $\sigma _{e}$ only participate in the charged lepton Yukawa interactions and whose inclusion is crucial to get a diagonal charged lepton mass matrix. On the other hand, the remaining gauge singlet scalars, i.e., $\varphi $, $\phi $, $\eta $, $\chi $ and $\xi $ only appear in the neutrino Yukawa terms, which yield a viable and very predictive mass matrix for light active neutrinos. Thus, the inclusion of these
scalar fields is necessary to have a highly predictive model for the lepton
sector with only two effective parameters in the light active neutrino
sector that allows to successfully reproduce the six experimental values of
the physical observables of the neutrino sector, i.e., the two neutrino mass
squared splittings, the three leptonic mixing angles and the leptonic Dirac
CP violating phase.

We additionally introduce several driving fields, which are grouped into
eleven $S_{4}$ singlets, i.e., $X_{k}$ ($k=1,2,\cdots ,11$) and four $S_{4}$
triplets, i.e., $\Phi $, $\Delta $, $\Theta $ and $\Xi $. These driving
fields are crucial for determining the vacuum aligments of the $S_{4}$
doublets and triplets in our model (to be specified below) that give rise to
a diagonal SM charged lepton mass matrix and to a highly predictive
and viable light active neutrino mass matrix, having only two free effective
parameters.

In our model, the SM gauge symmetry is supplemented by the inclusion of the $%
S_{4}\times U\left( 1\right) \times Z_{3}\times Z_{6}\times Z_{9}\times
U\left( 1\right) _{R}$ symmetry. We choose $S_{4}$ since it is the smallest
non abelian group having doublet, triplet and singlet irreducible
representations, thus allowing us to naturally accommodate the three
families of the SM left handed leptonic fields into a $S_{4}$ triplet and
the four gauge singlet right handed Majorana neutrinos into two $S_{4}$ singlets and one $S_{4}$ 
doublet, which is crucial to have highly predictive model that successfully
describes lepton masses and mixings. The $Z_{3}\times Z_{6}$ discrete
symmetry allows to get a diagonal SM charged lepton mass matrix and Dirac
neutrino mass matrix that yields a predictive and viable light active
neutrino mass matrix. Thus, the leptonic mixing in our model arises
from the neutrino sector. The $Z_{9}$ discrete symmetry sets the SM charged
lepton mass hierarchy. It is worth mentioning that despite its extended particle spectrum and symmetries, each introduced field and symmetry plays its own
role (described above) in predicting viable textures for the lepton sector that allows to successfully reproduce the experimental values of the six physical neutrino sector observables with only two effective parameters in the light active neutrino sector. This is achieved without the need to introduce hierarchy between the Yukawa couplings. Furthermore, the spontaneous breaking of the $%
S_{4}\times Z_{3}\times Z_{6}\times Z_{9}$ at very high energy, gives rise
to the SM charged lepton mass hierarchy. Besides that, the spontaneous
breaking of the $U\left( 1\right) $ global symmetry, which is assumed to
take place at the TeV scale is crucial to generate a renormalizable and a
non renormalizable mass terms involving gauge singlet right handed Majorana
neutrinos, required for the implementation of the inverse seesaw mechanism
that produce small masses for light active neutrinos. Note we have
introduced a $U\left( 1\right) _{R}$ symmetry under which the chiral
supermultiplets containing the SM fermions have charge equal $+1$, whereas
the driving fields $X_{k}$ ($k=1,2,\cdots ,11$), $\Phi $, $\Delta $, $\Theta 
$ and $\Xi $ have $U\left( 1\right) _{R}$ charge equal to $+2$ and the
remaining scalar fields are neutral under this symmetry. As a consequence of
that $U\left( 1\right) _{R}$ charge assignment, the aforementioned driving
fields can only appear linearly in the superpotential and do not feature
Yukawa interactions with SM fermions. The inclusion of the aforementioned
driving files, whose corresponding superpotential is given in Appendix \ref%
{WS4}, is necessary for achiving the following 
VEV configuration of the $S_{4}$ doublets and triplet scalars in our model: 
\begin{table}[th]
\resizebox{18cm}{!}{
\renewcommand{\arraystretch}{1.2}
\centering%
\begin{tabular}{|c|c|c|c|c|c|c|c|c|c|c|c|c|c|c|c|c|c|c|c|c|c|c|c|c|c|c|c|c|c|c|c|c|c|}
\hline\hline
& $l_{L}$ & $l_{1R}$ & $l_{2R}$ & $l_{3R}$ & $\nu _{1R}$ & $\nu _{2R}$ & $%
N_{R}$ & $H_{u}$ & $H_{d}$ & $\rho $ & $\varphi $ & $\phi $ & $\eta $ & $%
\sigma _{\mu }$ & $\sigma _{\tau }$ & $\sigma _{e}$ & $\chi $ & $\xi $ & $%
X_{1}$ & $X_{2}$ & $X_{3}$ & $X_{4}$ & $X_{5}$ & $X_{6}$ & $X_{7}$ & $X_{8}$
& $X_{9}$ & $X_{10}$ & $X_{11}$ & $\Phi $ & $\Delta $ & $\Theta $ & $\Xi $
\\ \hline
$S_{4}$ & $\mathbf{3}$ & $\mathbf{\mathbf{1}^{\prime }}$ & $\mathbf{1}$ & $%
\mathbf{1}$ & $\mathbf{1}$ & $\mathbf{1}$ & $\mathbf{2}$ & $\mathbf{1}$ & $%
\mathbf{1}$ & $\mathbf{1}$ & $\mathbf{2}$ & $\mathbf{2}$ & $\mathbf{\mathbf{2%
}}$ & $\mathbf{3}$ & $\mathbf{3}$ & $\mathbf{3}$ & $\mathbf{3}$ & $\mathbf{3}
$ & $\mathbf{\mathbf{1}}$ & $\mathbf{\mathbf{1}}$ & $\mathbf{\mathbf{1}}$ & $%
\mathbf{\mathbf{1}}$ & $\mathbf{\mathbf{1}}$ & $\mathbf{1^{\prime }}$ & $\mathbf{1}$ & $%
\mathbf{\mathbf{1}}$ & $\mathbf{1}^{\prime }$ & $\mathbf{\mathbf{1}}$ & $%
\mathbf{\mathbf{1}}$ & $\mathbf{3}$ & $\mathbf{3}$ & $\mathbf{3}$ & $\mathbf{%
3}^{\prime }$ \\ \hline
$U\left( 1\right) $ & $1$ & $2$ & $2$ & $2$ & $1$ & $1$ & $-2$ & $0$ & $-1$
& $0$ & $-3$ & $-1$ & $-1$ & $0$ & $0$ & $0$ & $0$ & $0$ & $2$ & $2$ & $0$ & 
$0$ & $1$ & $3$ & $0$ & $0$ & $9$ & $3$ & $3$ & $0$ & $0$ & $0$ & $0$ \\ 
\hline
$Z_{3}$ & $0$ & $-1$ & $-1$ & $0$ & $0$ & $-1$ & $0$ & $0$ & $0$ & $0$ & $0$
& $0$ & $-1$ & $1$ & $0$ & $1$ & $0$ & $1$ & $2$ & $0$ & $-1$ & $-1$ & $-2$
& $2$ & $-1$ & $0$ & $0$ & $-2$ & $-2$ & $-2$ & $0$ & $-2$ & $-1$ \\ \hline
$Z_{6}$ & $0$ & $3$ & $1$ & $1$ & $0$ & $0$ & $0$ & $0$ & $1$ & $0$ & $-1$ & 
$0$ & $0$ & $-2$ & $-2$ & $-4$ & $0$ & $0$ & $0$ & $0$ & $4$ & $0$ & $0$ & $%
0 $ & $2$ & $2$ & $3$ & $-1$ & $1$ & $4$ & $4$ & $2$ & $4$ \\ \hline
$Z_{9}$ & $-4$ & $4$ & $0$ & $-2$ & $0$ & $0$ & $0$ & $0$ & $0$ & $-1$ & $0$
& $0$ & $0$ & $0$ & $0$ & $0$ & $0$ & $0$ & $0$ & $0$ & $0$ & $0$ & $0$ & $0$
& $0$ & $0$ & $0$ & $0$ & $0$ & $0$ & $0$ & $0$ & $0$ \\ \hline
$U(1)_{R}$ & $1$ & $1$ & $1$ & $1$ & $1$ & $1$ & $1$ & $0$ & $0$ & $0$ & $0$
& $0$ & $0$ & $0$ & $0$ & $0$ & $0$ & $0$ & $2$ & $2$ & $2$ & $2$ & $2$ & $2$
& $2$ & $2$ & $2$ & $2$ & $2$ & $2$ & $2$ & $2$ & $2$ \\ \hline
\end{tabular}}%
\caption{Leptonic and scalar field assigments under the $S_{4}\times U\left(
1\right) \times Z_{3}\times Z_{6}\times Z_{9}\times U\left( 1\right) _{R}$
symmetry.}
\label{Themodel}
\end{table}
\begin{eqnarray}
\left\langle \varphi \right\rangle &=&v_{\varphi }\left( 1,\omega \right) ,%
\hspace{1cm}\left\langle \phi \right\rangle =v_{\phi }\left( 0,1\right) ,%
\hspace{1cm}\left\langle \eta \right\rangle =v_{\eta }\left( 1,0\right) ,%
\hspace{1cm}\left\langle \chi \right\rangle =v_{\chi }\left( 0,1,1\right) , 
\notag \\
\left\langle \xi \right\rangle &=&v_{\xi }\left( 1,3,1\right) ,\hspace{1cm}%
\left\langle \sigma _{\mu }\right\rangle =v_{\sigma _{\mu }}\left(
0,1,0\right) ,\hspace{1cm}\left\langle \sigma _{\tau }\right\rangle
=v_{\sigma _{\tau }}\left( 0,0,1\right) ,\hspace{1cm}\left\langle \sigma
_{e}\right\rangle =v_{\sigma _{e}}\left( 1,0,0\right) ,  \label{VEV}
\end{eqnarray}%
where $\omega =e^{\frac{2\pi i}{3}}$.

Since the spontaneous breaking of the $S_{4}\times Z_{3}\times Z_{6}\times
Z_{9}$ discrete group gives rise to the hierarchy of charged lepton masses,
we set the vacuum expectation values (VEVs) of the different gauge singlet
scalars with respect to the Wolfenstein parameter $\lambda =0.225$ and the
model cutoff $\Lambda $, as follows: 
\begin{equation}
v_{\phi }\sim v_{\eta }\sim v_{\varphi }\sim \mathcal{O}(1)\mathrm{TeV}%
<<v_{\chi }\sim v_{\xi }\sim v_{\sigma _{e}}\sim v_{\sigma _{\mu }}\sim
v_{\sigma _{\tau }}\sim v_{\rho }\sim \lambda \Lambda .  \label{VEVhierarchy}
\end{equation}%
Here, for the sake of simplicity, the VEVs $v_{\varphi }$, $v_{\phi }$, $%
v_{\eta }$, $v_{\rho }$, $v_{\chi }$, $v_{\xi }$, $v_{\sigma _{\mu }}$ and $%
v_{\sigma _{\tau }}$ are assumed to be real. As it will be shown in section \ref{leptonmassesandmixings}, the assumption of Eq. (\ref{VEVhierarchy}) will allow to explain the SM charged lepton mass hierarchy since it will relate the SM charged lepton masses with different powers of the Wolfenstein parameter times $\mathcal{O}(1)$ coefficients. It is worth mentioning that the model cutoff scale can be interpreted as the scale of the UV completion of the model, e.g. the masses Froggatt-Nielsen messenger fields. Furthermore, notice that the gauge singlet scalar fields $\phi$, $\eta$ and $\varphi$ are assumed to get VEVs at the TeV scale in order to have TeV scale sterile neutrinos, thus allowing to have a model testable at colliders. Thus, the hierarchy in the VEVs of the gauge singlet scalar fields shown in Eq. (\ref{VEVhierarchy}) is motivated in order to have TeV scale sterile neutrinos and to explain the SM charged lepton mass hierarchy. Such two scale VEV hierarchy can be explained by having appropiate relations between the different mass coefficients of the bilinear terms of the scalar potential and the VEVs of such scalar fields. 
To show this explicitly, we consider the simplified case of two singlet scalar fields $S_1$ and $S_2$, whose VEVs satisfy the hierarchy $v_{S_{2}}>>v_{S_{1}}$. The corresponding scalar potential involving such fields takes the form:
\begin{equation}
V=-\mu _{S_{1}}^{2}\left\vert S_{1}\right\vert ^{2}-\mu
_{S_{2}}^{2}\left\vert S_{2}\right\vert ^{2}+\lambda _{1}\left\vert
S_{1}\right\vert ^{4}+\lambda _{2}\left\vert S_{2}\right\vert ^{4}+\lambda
_{3}\left\vert S_{1}\right\vert ^{2}\left\vert S_{2}\right\vert ^{2}.
\end{equation}
Its minimization yields the following relations:
\begin{equation}
\mu _{S_{1}}^{2}=2\lambda _{1}v_{S_{1}}^{2}+\lambda _{3}v_{S_{2}}^{2},%
\hspace{1.5cm}\mu _{S_{2}}^{2}=2\lambda _{2}v_{S_{2}}^{2}+\lambda
_{3}v_{S_{1}}^{2}.
\end{equation}
Consequently, the VEV hierarchy $v_{S_{2}}>>v_{S_{1}}$, can be justified by requiring $\mu _{S_{2}}^{2}\simeq 2\mu _{S_{1}}^{2}$ and considering the case where the quartic scalar couplings satisfy $\lambda_i\simeq\lambda$ ($i=1,2,3$). A straightforward but tedious extension of the aforementioned argument will give rise to large a set of relations between the different mass coefficients of the bilinear terms of the scalar potential and the VEVs of the large number of gauge singlet scalar fields of our model that will yield the VEV hierarchy shown in Eq. \ref{VEVhierarchy}.

With the above particle content, we have the following relevant charged
lepton and neutrino Yukawa terms: 
\begin{equation}
-\mathcal{L}_{Y}^{\left( l\right) }=y_{1}^{\left( l\right) }\left( \overline{%
l}_{L}H_{d}\sigma _{e}\right) _{\mathbf{\mathbf{1}^{\prime }}}l_{1R}\frac{%
\rho ^{8}}{\Lambda ^{9}}+y_{2}^{\left( l\right) }\left( \overline{l}%
_{L}H_{d}\sigma _{\mu }\right) _{\mathbf{\mathbf{1}}}l_{2R}\frac{\rho ^{4}}{%
\Lambda ^{5}}+y_{3}^{\left( l\right) }\left( \overline{l}_{L}H_{d}\sigma
_{\tau }\right) _{\mathbf{\mathbf{1}}}l_{3R}\frac{\rho ^{2}}{\Lambda ^{3}}%
+H.c  \label{Lyl}
\end{equation}%
\begin{eqnarray}
-\mathcal{L}_{Y}^{\left( \nu \right) } &=&y_{1}^{\left( \nu \right) }\left( 
\overline{l}_{L}H_{u}\chi \right) _{\mathbf{\mathbf{1}}}\nu _{1R}\frac{\rho
^{4}}{\Lambda ^{5}}+y_{2}^{\left( \nu \right) }\left( \overline{l}%
_{L}H_{u}\xi \right) _{\mathbf{\mathbf{1}}}\nu _{2R}\frac{\rho ^{4}}{\Lambda
^{5}}  \label{Lynu} \\
&&+y_{1\nu N}\overline{\nu }_{1R}\left( \phi N_{R}^{C}\right) _{\mathbf{1}}%
\frac{1}{\Lambda }+y_{2\nu N}\overline{\nu }_{2R}\left( \eta
N_{R}^{C}\right) _{\mathbf{1}}+y_{N}\left( \overline{N}_{R}N_{R}^{C}\right)
_{\mathbf{2}}\varphi \frac{H_{u}H_{d}}{\Lambda ^{2}}+H.c  \notag
\end{eqnarray}

\section{Lepton masses and mixings}
\label{leptonmassesandmixings} 
From the charged lepton Yukawa terms, we find
that the charged lepton mass matrix is diagonal and the SM charged lepton
masses are given by: 
\begin{equation}
m_{e}=y_{1}^{\left( l\right) }\frac{v_{\sigma _{e}}v_{\rho }^{8}}{\sqrt{2}%
\Lambda ^{9}}v_{H_{d}}=a_{1}^{\left( l\right) }\lambda ^{9}\frac{v}{\sqrt{2}}%
,\hspace{1cm}m_{\mu }=y_{2}^{\left( l\right) }\frac{v_{\sigma _{\mu
}}v_{\rho }^{4}}{\sqrt{2}\Lambda ^{5}}v_{H_{d}}=a_{2}^{\left( l\right)
}\lambda ^{5}\frac{v}{\sqrt{2}},\hspace{1cm}m_{\tau }=y_{3}^{\left( l\right)
}\frac{v_{\sigma _{\tau }}v_{\rho }^{2}}{\sqrt{2}\Lambda ^{3}}%
v_{H_{d}}=a_{3}^{\left( l\right) }\lambda ^{3}\frac{v}{\sqrt{2}},
\end{equation}%
where $a_{1}^{\left( l\right) }$, $a_{2}^{\left( l\right) }$ and $%
a_{3}^{\left( l\right) }$ are real $\mathcal{O}(1)$ dimensionless parameters and we have assumed that $v_{H_{d}}\sim v/\sqrt{2}$, being $v=246$ GeV the electroweak symmetry breaking scale.

Regarding the neutrino sector, from the Eq. (\ref{Lynu}), we find the
following neutrino mass terms: 
\begin{equation}
-\mathcal{L}_{mass}^{\left( \nu \right) }=\frac{1}{2}\left( 
\begin{array}{ccc}
\overline{\nu _{L}^{C}} & \overline{\nu _{R}} & \overline{N_{R}}%
\end{array}%
\right) M_{\nu }\left( 
\begin{array}{c}
\nu _{L} \\ 
\nu _{R}^{C} \\ 
N_{R}^{C}%
\end{array}%
\right) +H.c,  \label{Lnu}
\end{equation}%
where the neutrino mass matrix is given by: 
\begin{equation}
M_{\nu }=\left( 
\begin{array}{ccc}
0_{3\times 3} & m_{D} & 0_{3\times 2} \\ 
m_{D}^{T} & 0_{2\times 2} & M \\ 
0_{2\times 3} & M^{T} & \mu%
\end{array}%
\right) ,  \label{Mnu}
\end{equation}%
%
%
%
%
%
%
%
%
%
%
%
%
%
%
%
%
%
%
where $0_{n\times m}$ are $n\times m$ dimensional submatrices consisting of
all zeroes and the other submatrices in the flavour basis have the
structure: 
\begin{eqnarray}
m_{D} &=&v_{H_{u}}\left( 
\begin{array}{cc}
0 & b \\ 
a & 3b \\ 
a & b%
\end{array}%
\right) ,\hspace{0.7cm}\hspace{0.7cm}M=m_{N}\left( 
\begin{array}{cc}
1 & 0 \\ 
0 & z%
\end{array}%
\right) ,\hspace{0.7cm}\hspace{0.7cm}\mu =\frac{y_{N}v_{H_{u}}v_{H_{d}}v_{%
\varphi }}{\Lambda ^{2}}\left( 
\begin{array}{cc}
1 & 0 \\ 
0 & \omega%
\end{array}%
\right) ,\hspace{0.7cm}\hspace{0.7cm}\omega =e^{\frac{2\pi i}{3}},  \notag \\
a &=&y_{1}^{\left( \nu \right) }\frac{v_{\chi }v_{\rho }^{4}}{\Lambda ^{5}}%
=x_{1}^{\left( \nu \right) }\lambda ^{5},\hspace{0.7cm}\hspace{0.7cm}%
b=y_{2}^{\left( \nu \right) }\frac{v_{\xi }v_{\rho }^{4}}{\Lambda ^{5}}%
=x_{2}^{\left( \nu \right) }\lambda ^{5},\hspace{0.7cm}\hspace{0.7cm}%
m_{N}=y_{1\nu N}v_{\phi },\hspace{0.7cm}\hspace{0.7cm}z=y_{2\nu N}\frac{%
v_{\eta }}{v_{\phi }}.  \label{Mnublocks}
\end{eqnarray}%
The above mass matrices in Eqs. (\ref{Mnu}), (\ref{Mnublocks}) have precisely the
desired LIS structure given in Eqs. (\ref{Mnu0}), (\ref{Mnublocks0}) in
Section~\ref{Intro}.

As shown in detail in Ref. \cite{Catano:2012kw}, the full rotation matrix
that diagonalizes a neutrino mass matrix of the form of Eq. (\ref{Mnu}) is
given by: 
\begin{equation}
\mathbb{R}=%
\begin{pmatrix}
R_{\nu } & R_{1}R_{M}^{\left( 1\right) } & R_{2}R_{M}^{\left( 2\right) } \\ 
-\frac{(R_{1}^{\dagger }+R_{2}^{\dagger })}{\sqrt{2}}R_{\nu } & \frac{(1-S)}{%
\sqrt{2}}R_{M}^{\left( 1\right) } & \frac{(1+S)}{\sqrt{2}}R_{M}^{\left(
2\right) } \\ 
-\frac{(R_{1}^{\dagger }-R_{2}^{\dagger })}{\sqrt{2}}R_{\nu } & \frac{(-1-S)%
}{\sqrt{2}}R_{M}^{\left( 1\right) } & \frac{(1-S)}{\sqrt{2}}R_{M}^{\left(
2\right) }%
\end{pmatrix}%
,  \label{U}
\end{equation}%
where 
\begin{equation}
S=-\frac{1}{4}M^{-1}\mu ,\hspace{1cm}\hspace{1cm}R_{1}\simeq R_{2}\simeq 
\frac{1}{\sqrt{2}}m_{D}^{\ast }M^{-1}=\frac{1}{\sqrt{2}m_{N}}m_{D}^{\ast }.
\end{equation}

The light active masses arise from an inverse seesaw mechanism and the
physical neutrino mass matrices are: 
\begin{equation}
m_{\nu }=m_{D}\left( M^{T}\right) ^{-1}\mu M^{-1}m_{D}^{T},\hspace{0.7cm}%
M_{\nu }^{\left( 1\right) }=-\frac{1}{2}\left( M+M^{T}\right) +\frac{1}{2}%
\mu ,\hspace{0.7cm}M_{\nu }^{\left( 2\right) }=\frac{1}{2}\left(
M+M^{T}\right) +\frac{1}{2}\mu ,  \label{M1nu}
\end{equation}

where $m_{\nu }$ corresponds to the active neutrino mass matrix whereas $%
M_{\nu }^{\left( 1\right) }$ and $M_{\nu }^{\left( 2\right) }$ are the
exotic neutrino mass matrices.

Note that the physical neutrino spectrum is composed of three light active
neutrinos and four exotic neutrinos. The exotic neutrinos are pseudo-Dirac,
with masses $\sim \pm \frac{1}{2}\left(M+M^{T}\right) $ and a small
splitting $\mu $. Furthermore, $R_{\nu }$, $R_{M}^{\left( 1\right) }$ and $%
R_{M}^{\left( 2\right) }$ are the rotation matrices which diagonalize $%
m_{\nu }$, $M_{\nu }^{\left( 1\right) }$ and $M_{\nu }^{\left( 2\right) }$,
respectively. Since in our model $M_{\nu }^{\left( 1\right) }$ and $M_{\nu
}^{\left( 2\right) }$ are diagonal, $R_{M}^{\left( 1\right) }$ and $%
R_{M}^{\left( 2\right) }$ are equal to the $2\times 2$ identity matrix, the
rotation matrix $\mathbb{R}$ of Eq. (\ref{U}) can be rewritten as follows: 
\begin{equation}
\mathbb{R}=%
\begin{pmatrix}
R_{\nu } & R_{1} & R_{2} \\ 
-\frac{(B_{2}^{\dagger }+B_{3}^{\dagger })}{\sqrt{2}}R_{\nu } & \frac{(1-S)}{%
\sqrt{2}} & \frac{(1+S)}{\sqrt{2}} \\ 
-\frac{(B_{2}^{\dagger }-B_{3}^{\dagger })}{\sqrt{2}}R_{\nu } & \frac{(-1-S)%
}{\sqrt{2}} & \frac{(1-S)}{\sqrt{2}}%
\end{pmatrix}%
.
\end{equation}

Furthermore, using Eq. (\ref{U}) we find that the neutrino fields $\nu
_{L}=\left( \nu _{1L},\nu _{2L},\nu _{3L}\right) ^{T}$, $\nu _{R}^{C}=\left(
\nu _{1R}^{C},\nu _{2R}^{C}\right) $ and $N_{R}^{C}=\left(
N_{1R}^{C},N_{2R}^{C}\right) $ are related with the neutrino mass
eigenstates by the following relations: 
\begin{equation}
\left( 
\begin{array}{c}
\nu _{L} \\ 
\nu _{R}^{C} \\ 
N_{R}^{C}%
\end{array}%
\right) =\mathbb{R}\Omega _{L}\simeq 
\begin{pmatrix}
R_{\nu } & R_{1} & R_{2} \\ 
-\frac{(B_{2}^{\dagger }+B_{3}^{\dagger })}{\sqrt{2}}R_{\nu } & \frac{(1-S)}{%
\sqrt{2}} & \frac{(1+S)}{\sqrt{2}} \\ 
-\frac{(B_{2}^{\dagger }-B_{3}^{\dagger })}{\sqrt{2}}R_{\nu } & \frac{(-1-S)%
}{\sqrt{2}} & \frac{(1-S)}{\sqrt{2}}%
\end{pmatrix}%
\left( 
\begin{array}{c}
\Omega _{L}^{\left( 1\right) } \\ 
\Omega _{L}^{\left( 2\right) } \\ 
\Omega _{L}^{\left( 3\right) }%
\end{array}%
\right) ,\hspace{0.5cm}\hspace{0.5cm}\hspace{0.5cm}\hspace{0.5cm}\Omega
_{L}=\left( 
\begin{array}{c}
\Omega _{L}^{\left( 1\right) } \\ 
\Omega _{L}^{\left( 2\right) } \\ 
\Omega _{L}^{\left( 3\right) }%
\end{array}%
\right) ,
\end{equation}%
where $\Omega _{jL}^{\left( 1\right) }$ ($j=1,2,3$), $\Omega _{kL}^{\left(
2\right) }$ and $\Omega _{kL}^{\left( 3\right) }$ ($k=1,2$) are the three
active neutrinos and four exotic neutrinos, respectively.

Using Eq. (\ref{M1nu}), the light active neutrino mass matrix arising from
the inverse seesaw mechanism takes the form: 
\begin{equation}
m_{\nu }=m_{\nu a}\left( 
\begin{array}{ccc}
0 & 0 & 0 \\ 
0 & 1 & 1 \\ 
0 & 1 & 1%
\end{array}%
\right) +m_{\nu b}\omega \left( 
\begin{array}{ccc}
1 & 3 & 1 \\ 
3 & 9 & 3 \\ 
1 & 3 & 1%
\end{array}%
\right) ,\hspace{0.7cm}\hspace{0.7cm}m_{\nu a}=\frac{%
a^{2}y_{N}v_{H_{u}}^{3}v_{H_{d}}v_{\varphi }}{y_{1\nu N}^{2}v_{\phi
}^{2}\Lambda ^{2}},\hspace{0.7cm}\hspace{0.7cm}m_{\nu b}=\frac{%
b^{2}z^{2}y_{N}v_{H_{u}}^{3}v_{H_{d}}v_{\varphi }}{y_{1\nu N}^{2}v_{\phi
}^{2}\Lambda ^{2}}.  \label{mnu}
\end{equation}

The low energy neutrino mass matrix in Eq.~\ref{mnu} is of the highly
predictive LS form given in Eq.~(\ref{mnu0}) which gives a good fit to low
energy neutrino data using the parameter values discussed for example in~%
\cite{Ballett:2016yod}. The neutrino mass squared splittings, light active
neutrino masses, leptonic mixing angles and CP violating phase for the
scenario of normal neutrino mass hierarchy can be very well reproduced with
only two effective free parameters, whose values are given by~\cite%
{Ballett:2016yod}: 
\begin{equation}
m_{\nu a}\simeq 26.57\mbox{meV},\hspace{1cm}m_{\nu b}\simeq 2.684\,\mbox{meV}%
.  \label{bestfitpoint}
\end{equation}%
Thus, using the numerical value for $m_{\nu b}$ given by Eq. (\ref%
{bestfitpoint}) and considering $v_{H_{u}}\sim v_{H_{d}}\sim \frac{v}{\sqrt{2%
}}\sim 174$ GeV, $v_{\phi }\sim v_{\varphi }\sim 1$ TeV, $y_{\nu N}\sim
y_{N}\sim 1$, $b\sim \lambda ^{5}$, with $\lambda =0.225$ and $v=246$ GeV,
we estimate our model cutoff as $\Lambda \sim 3\times 10^{5}$ GeV, in order
to naturally reproduce the smallness of the light active neutrino masses.

In addition, we find that the light active neutrino masses are: 
\begin{equation}
m_{1}=0,\hspace{1cm}m_{2}=8.59\mbox{meV}\hspace{1cm}m_{3}=49.81\mbox{meV}.
\end{equation}%
\begin{table}[tbh]
\begin{center}
\begin{tabular}{c|l|l|l|l|l}
\hline\hline
Observable & Model & bpf $\pm 1\sigma $ \cite{deSalas:2017kay} & bpf $\pm
1\sigma $ \cite{Esteban:2018azc} & $3\sigma $ range \cite{deSalas:2017kay} & 
$3\sigma $ range \cite{Esteban:2018azc} \\ \hline
$\Delta m_{21}^{2}$ [$10^{-5}$eV$^{2}$] & \quad $7.38$ & \quad $%
7.55_{-0.16}^{+0.20}$ & \quad $7.39_{-0.20}^{+0.21}$ & \quad $7.05-8.14$ & 
\quad $6.79-8.01$ \\ \hline
$\Delta m_{31}^{2}$ [$10^{-3}$eV$^{2}$] & \quad $2.48$ & \quad $2.50\pm 0.03$
& \quad $2.525_{-0.031}^{+0.033}$ & \quad $2.41-2.60$ & \quad $2.431-2.622$
\\ \hline
$\theta _{12}^{(l)}(^{\circ })$ & \quad $34.32$ & \quad $34.5_{-1.0}^{+1.2}$
& \quad $33.82_{-0.76}^{+0.78}$ & \quad $31.5-38.0$ & \quad $31.61-36.27$ \\ 
\hline
$\theta _{13}^{(l)}(^{\circ })$ & \quad $8.67$ & \quad $8.45_{-0.14}^{+0.16}$
& \quad $8.61_{-0.13}^{+0.12}$ & \quad $8.0-8.9$ & \quad $8.22-8.98$ \\ 
\hline
$\theta _{23}^{(l)}(^{\circ })$ & \quad $45.77$ & \quad $47.9_{-1.7}^{+1.0}$
& \quad $49.7_{-1.1}^{+0.9}$ & \quad $41.8-50.7$ & \quad $40.9-52.2$ \\ 
\hline
$\delta _{CP}^{(l)}(^{\circ })$ & $-86.67$ & \quad $-142_{-27}^{+38}$ & 
\quad $217_{-28}^{+40}$ & \quad $157-349$ & \quad $135-366$ \\ \hline\hline
\end{tabular}%
%
%
%
%
%
%
%
%
%
%
%
%
%
%
%
%
%
%
%
%
%
%
%
%
%
%
%
%
%
%
%
%
%
%
%
%
%
%
%
%
%
%
\end{center}
\caption{Model and experimental values of the light active neutrino masses,
leptonic mixing angles and CP violating phase for the scenario of normal
(NH) neutrino mass hierarchy. The experimental values are taken from Refs. 
\protect\cite{deSalas:2017kay,Esteban:2018azc}}
\label{Neutrinos}
\end{table}
From Table \ref{Neutrinos}, it follows that the neutrino mass squared
splittings, i.e, $\Delta m_{21}^{2}$ and $\Delta m_{31}^{2}$, the leptonic
mixing angles $\theta _{12}^{(l)}$, $\theta _{23}^{(l)}$, $\theta
_{13}^{(l)} $ and the Dirac leptonic CP violating phase are consistent with
neutrino oscillation experimental data for the scenario of normal neutrino
mass hierarchy. It is remarkable that our model relies on only two effective
parameters in the light active neutrino sector that allows to successfully
reproduce six neutrino physical observables: neutrino mass squared
splittings, leptonic mixing angles and Dirac leptonic CP violating phase.
Let us note that, for the inverted neutrino mass hierarchy, the obtained
leptonic mixing parameters are very much outside the $3\sigma $
experimentally allowed range. Consequently, our model is only viable for the
scenario of normal neutrino mass hierarchy.

\section{Charged lepton flavor violating decays.}

\label{LFV} In this section we will discuss the implications of our model in
the lepton flavor violating decays $\mu \rightarrow e\gamma $, $\tau
\rightarrow \mu \gamma $ and $\tau \rightarrow e\gamma $. As mentioned in
the previous section, the physical sterile neutrino spectrum of our model is
composed of four TeV scale neutrinos, which are practically degenerate.
These heavy sterile neutrinos mix the active ones, with mixing angles of the
order of $\frac{1}{\sqrt{2}m_{N}}\left( m_{D}\right) _{in}$ \ ($i=1,2,3$ and 
$n=1,2$). The admixture of the heavy sterile neutrinos in the left-handed
charged current $SU_{2L}\times U_{1Y}$ weak interaction, gives rise to the $%
l_{i}\rightarrow l_{j}\gamma $ decay at one loop level, whose Branching
ratio takes the form \cite{Ilakovac:1994kj,Deppisch:2004fa,Lindner:2016bgg}: 
\begin{eqnarray}
Br\left( l_{i}\rightarrow l_{j}\gamma \right) &=&\frac{\alpha
_{W}^{3}s_{W}^{2}m_{l_{i}}^{5}}{256\pi ^{2}m_{W}^{4}\Gamma _{i}}\left\vert
G_{ij}\right\vert ^{2},\hspace{0.5cm}\hspace{0.5cm}\hspace{0.5cm}\hspace{%
0.5cm}  \notag \\
G_{ij} &=&\sum_{k}\left( \mathbb{R}^{\ast }\right) _{ik}\left( \mathbb{R}%
\right) _{jk}G_{\gamma }\left( \frac{m_{N_{k}}^{2}}{m_{W}^{2}}\right) \simeq
2\left( R_{1}R_{1}^{T}\right) _{ij}G_{\gamma }\left( \frac{m_{N}^{2}}{%
m_{W}^{2}}\right) =\frac{\left(m_{D}^{\ast }m_{D}^{\dagger }\right) _{ij}}{%
m_{N}^{2}}G_{\gamma }\left( \frac{m_{N}^{2}}{m_{W}^{2}}\right) ,  \notag \\
G_{\gamma } &=&-\frac{2x^{3}+5x^{2}-x}{4\left( 1-x\right) ^{2}}-\frac{3x^{3}%
}{2\left( 1-x\right) ^{4}}\ln x.  \label{Brmutoegamma}
\end{eqnarray}%
Thus, the Branching ratios for the $\mu \rightarrow e\gamma $, $\tau
\rightarrow \mu \gamma $ and $\tau \rightarrow e\gamma $ decays in our model
are respectively given by: 
\begin{eqnarray}
Br\left( \mu \rightarrow e\gamma \right) &=&\frac{9\alpha
_{W}^{3}s_{W}^{2}b^{4}v_{H_{u}}^{4}m_{\mu }^{5}}{256\pi ^{2}m_{W}^{4}\Gamma
_{\mu }m_{N}^{4}}\left\vert G_{\gamma }\left( \frac{m_{N}^{2}}{m_{W}^{2}}%
\right) \right\vert ^{2},  \notag \\
Br\left( \tau \rightarrow \mu \gamma \right) &=&\frac{%
9\alpha_{W}^{3}s_{W}^{2}b^{4}v_{H_{u}}^{4}m_{\tau }^{5}}{256\pi
^{2}m_{W}^{4}\Gamma _{\tau }m_{N}^{4}}\left\vert G_{\gamma }\left( \frac{%
m_{N}^{2}}{m_{W}^{2}}\right) \right\vert ^{2},  \notag \\
Br\left( \tau \rightarrow e\gamma \right) &=&\frac{\alpha
_{W}^{3}s_{W}^{2}b^{4}v_{H_{u}}^{4}m_{\tau }^{5}}{256\pi ^{2}m_{W}^{4}\Gamma
_{\tau }m_{N}^{4}}\left\vert G_{\gamma }\left( \frac{m_{N}^{2}}{m_{W}^{2}}%
\right) \right\vert ^{2},  \label{BrLFVsignals}
\end{eqnarray}%
being $\Gamma _{\mu }=3\times 10^{-19}$ GeV and $\Gamma _{\tau }=2.27\times
10^{-12}$ GeV the total muon and tau decay widths, respectively. 
Figure \ref{LFV} shows the allowed parameter space in the $m_{N}-bv_{H_{u}}$
and $m_{N}-x_{2}^{\left( \nu \right)}$ and $m_{N}-\tan\beta$ planes
consistent with the LFV constraints. The plots of Figure \ref{LFV} were
obtained by randomly generating the parameters $m_{N}$, $bv_{H_{u}}$ and $%
x_{2}^{\left( \nu \right) }$ (keeping in mind that $b=x_{2}^{\left( \nu
\right) }\lambda ^{5}$ (see Eq. (\ref{Mnublocks}))) in a range of values
where the Branching ratio for the $\mu \rightarrow e\gamma $ decay is below
its upper experimental limit of $4.2\times 10^{-13}$. To choose the region
where $bv_{H_{u}}$ was varied, we chose a scenario where $v_{H_{u}}=200$ GeV
with the dimensionless coupling $x_{2}^{\left( \nu \right) }$ in the range $%
1\lesssim x_{2}^{\left( \nu \right) }\lesssim \sqrt{4\pi }$, where the upper
bound of $\sqrt{4\pi }$ for $x_{2}^{\left( \nu \right) }$ corresponds to the
maximum value allowed by perturbativity. In what regards the third plot of
Figure \ref{LFV}, we have set $x_{2}^{\left( \nu \right) }$ equal to unity
and we have varied $\tan\beta=\frac{v_{H_{u}}}{v_{H_{d}}}$ in the range $5\lesssim \tan\beta\lesssim 50$
as done in Ref. \cite{Carena:2015uoe}.

As seen from Figure \ref{LFV}, the obtained values for the branching ratio
of the $\mu \rightarrow e\gamma $ decay are below its experimental upper
limit of $4.2\times 10^{-13}$ since these values are located in the range $8\times 10^{-14}\lesssim Br\left( \mu \rightarrow e\gamma \right) \lesssim
1.8\times 10^{-13}$, for a large region of parameter space of our model.
Furthermore, let us note that the branching ratio for the $\mu\to e\gamma$
decay has a low sensitivity with $\tan\beta$ when it is varied in the range $%
5\lesssim \tan\beta\lesssim 10$.

In the same region of parameter space, we found that the branching ratios
for the $\tau \rightarrow \mu \gamma $ and $\tau \rightarrow e\gamma $
decays are in the ranges $2\times 10^{-13}\lesssim Br\left( \tau \rightarrow
\mu \gamma \right) \lesssim 1.6\times 10^{-12}$ and $2\times
10^{-14}\lesssim Br\left( \tau \rightarrow e\gamma \right) \lesssim
1.8\times 10^{-13}$, respectively, which is well below their upper
experimental limits of $4.4\times 10^{-9}$ and $3.3\times 10^{-9}$,
respectively. Consequently, our model is highly consistent with the
constaints arising from lepton flavour violating decays for a large region
of parameter space. 
\begin{figure}[tbp]
\includegraphics[width=0.49\textwidth]{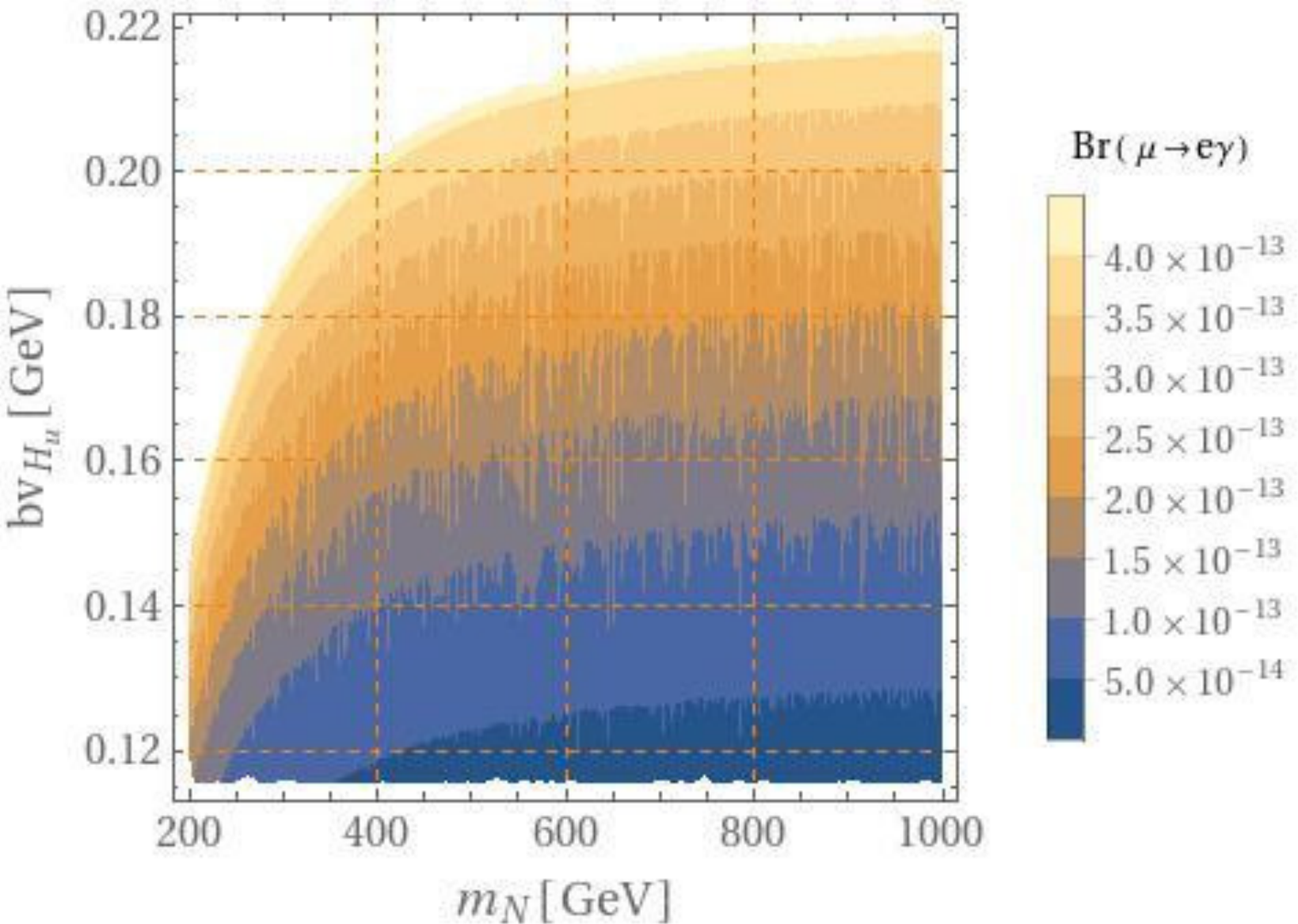}%
\includegraphics[width=0.49\textwidth]{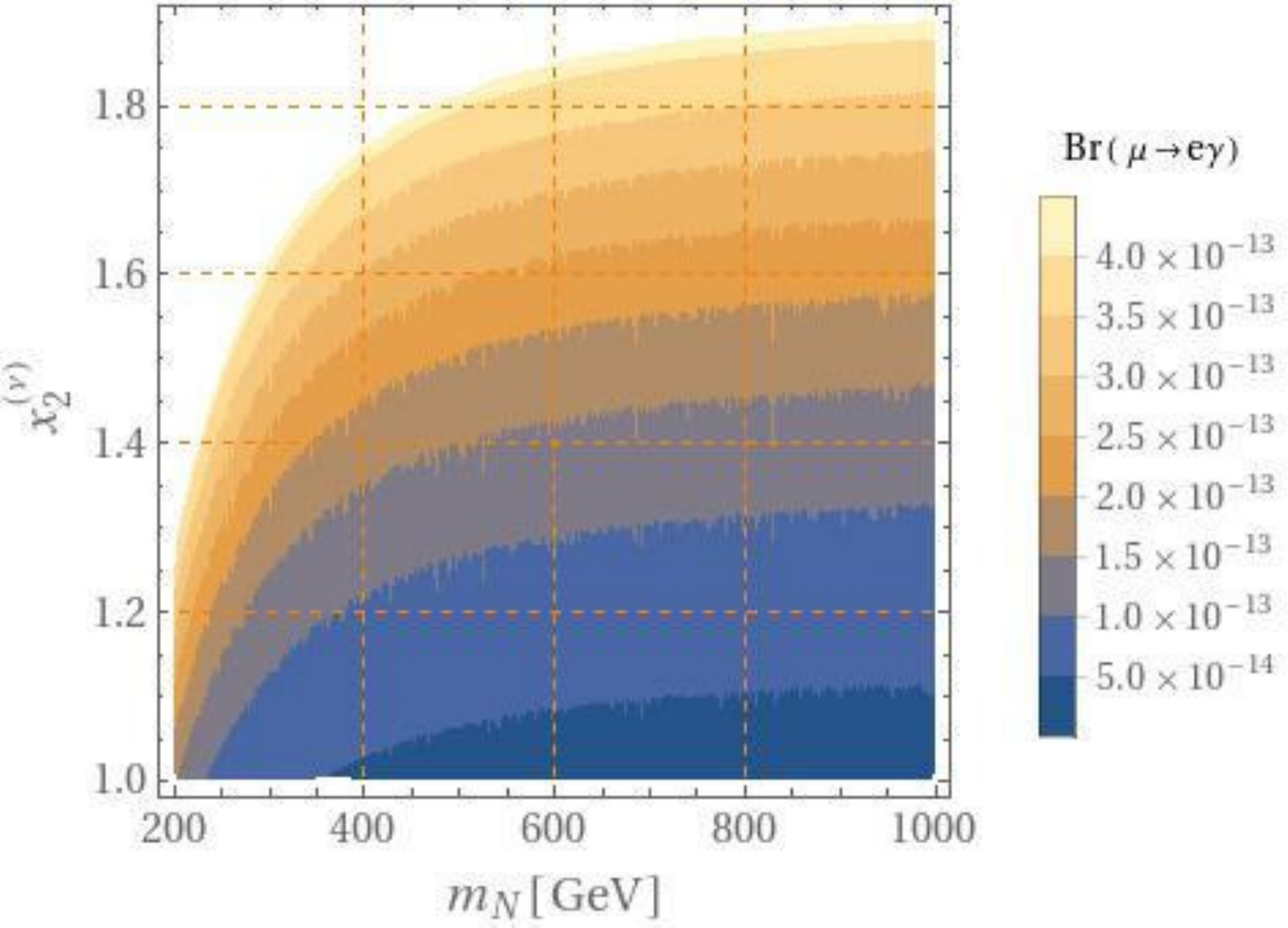} %
\includegraphics[width=0.49\textwidth]{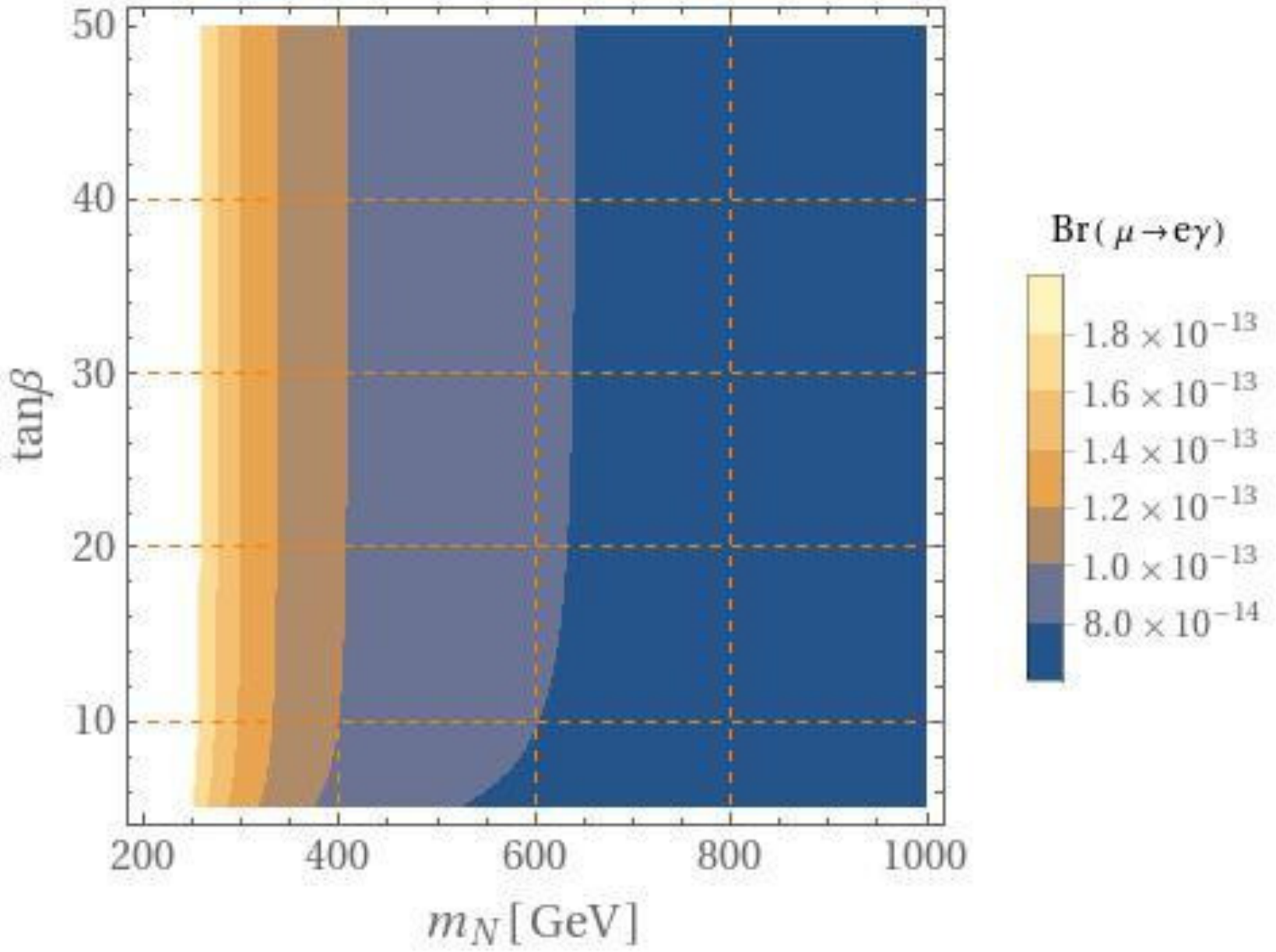}
\caption{Allowed parameter space in the $m_{N}-bv_{H_{u}}$, $%
m_{N}-x_{2}^{\left( \protect\nu \right) }$ and $m_{N}-\tan\protect\beta$
planes consistent with the LFV constraints. In the third plot $x_{2}^{\left( \protect\nu \right) }$ has been set equal to unity.}
\label{LFV}
\end{figure}
Given that future experiments such as Mu2e and COMET are expected to measure or bound lepton conversion in nuclei with much better precision than the radiative rare lepton decays, we proceed to determine the constraints imposed by lepton conversion in nuclei on the model parameter space. It is worth mentioning that the branching ratio for the $\mu^{-}-e^{-}$ conversion takes the form \cite{Lindner:2016bgg}:
\begin{equation}
CR\left(\mu-e\right)=\frac{\Gamma\left(\mu^{-}+Nucleus\left(A,Z\right)\rightarrow e^{-}+Nucleus\left(A,Z\right)\right)}{\Gamma\left(\mu^{-}+Nucleus\left(A,Z\right)\rightarrow\nu_{\mu}+Nucleus\left(A,Z-1\right)\right)}  
\end{equation}
Using an Effective Lagrangian approach for describing lepton flavor violating processes as done in \cite{Kuno:1999jp} and considering the low momentum limit where the off-shell contributions from photon exchange are negligible with respect to the contributions arising from real photon emision, the dipole operators dominate the conversion rate thus yielding the following relations \cite{Kuno:1999jp,Lindner:2016bgg}:
\begin{equation}
CR\left(\mu Ti\rightarrow eTi\right)\simeq\frac{1}{200}Br\left(\mu \rightarrow e\gamma\right)\hspace{1cm}CR\left(\mu Al\rightarrow eAl\right)\simeq\frac{1}{350}Br\left(\mu \rightarrow e\gamma\right)
\end{equation}
It is worth mentioning that the Effective field theory treatment used in \cite{Kuno:1999jp}, is valid for supersymmetric models like the one discussed in this paper.

\begin{figure}[tbp]
\includegraphics[width=0.47\textwidth]{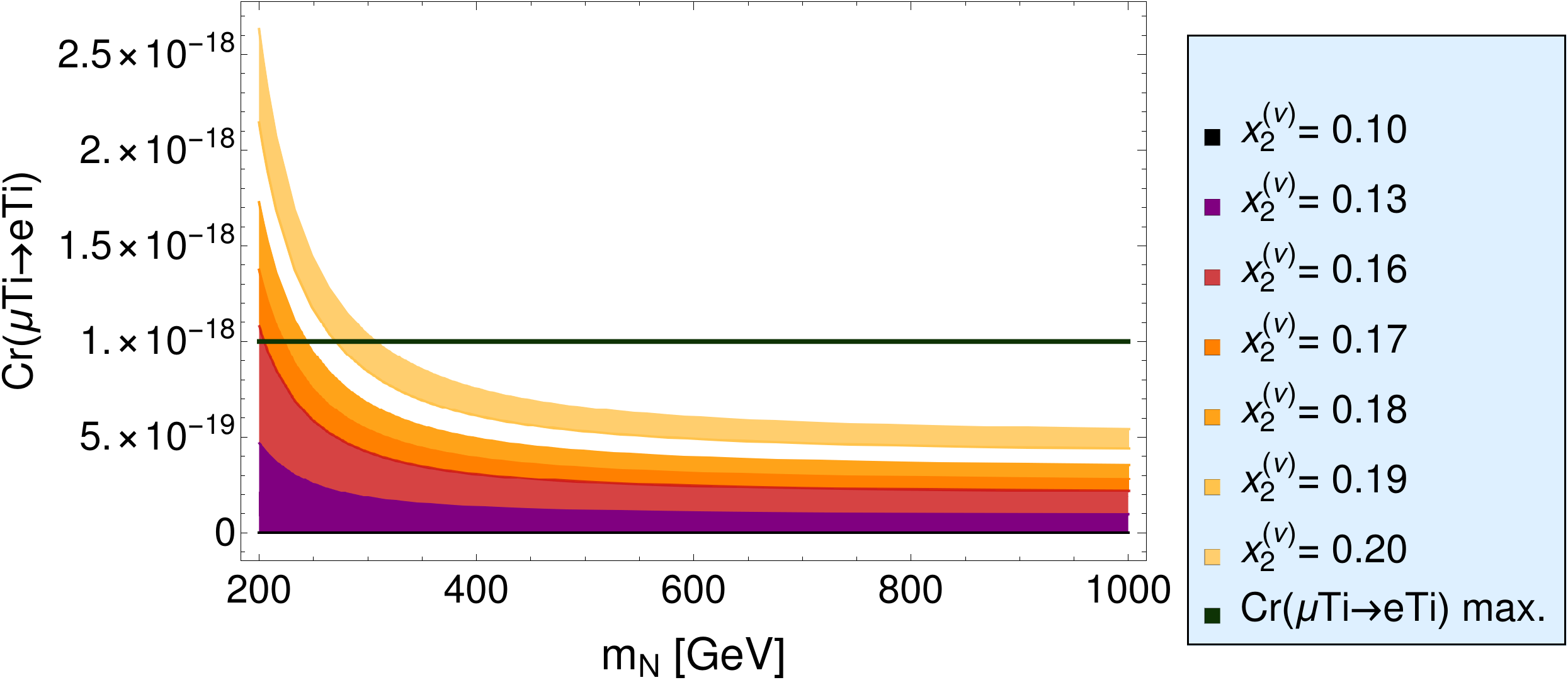}\includegraphics[width=0.47\textwidth]{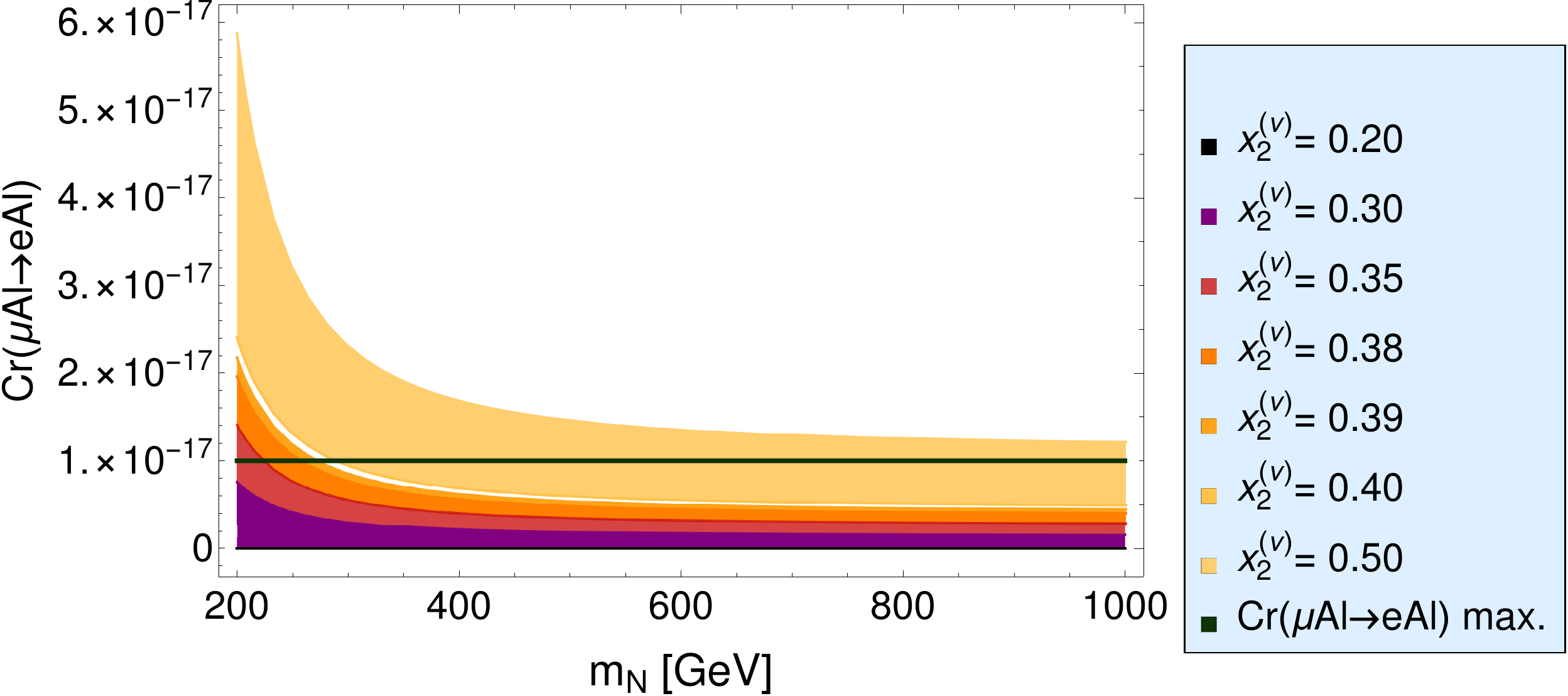}
\caption{$CR\left(\mu Ti\rightarrow eTi\right)$ (left plot) and $CR\left(\mu Al\rightarrow eAl\right)$ (right plot) as function of the sterile neutrino mass $m_N$ for different values of the dimensionless coupling $x_{2}^{\left( \nu \right) }$. The black horizontal line in each plot corresponds to the expected sensitivities of the next generation of experiments that will use Titanium \cite{Aysto:2001zs} and Aluminum \cite{Bernstein:2013hba} as targets, respectively. Here we have set $\tan\beta=5$.}
\label{CR}
\end{figure}

Figure \ref{CR} shows the $CR\left(\mu Ti\rightarrow eTi\right)$ (left plot) and $CR\left(\mu Al\rightarrow eAl\right)$ (right plot) parameters as function of the sterile neutrino mass $m_N$ for different values of the dimensionless coupling $x_{2}^{\left( \nu \right) }$. The black horizontal line in the left plot corresponds to the expected sensitivity of $\sim 10^{-18}$ of the CERN Neutrino Factory that will use Titanium as target \cite{Aysto:2001zs}. On the other hand, the black horizontal line in the right plot corresponds to the expected sensitivities of $\sim 10^{-17}$ of the next generation of experiments such as Mu2e and COMET \cite{Bernstein:2013hba}, where the Aluminum will be used as a target instead. In these plots we have set $\tan\beta=5$. These plots show that the next generation experiments where the Titanium and Aluminium will be used as targets, will rule out the part of the model parameter space where $x_{2}^{\left( \nu \right) }\gtrsim 0.2$ and $x_{2}^{\left( \nu \right) }\gtrsim 0.4$, respectively, for sterile neutrino masses larger than about $300$ GeV. Consequently, a precise measurement of lepton conversion in nuclei by future experiments will be crucial to set constraints on the active-sterile neutrino mixing angles, which will be crucial to determine the allowed region of parameter space of inverse seesaw models. 
Finally to close this section, it is worth mentioning that our results regarding the charged lepton flavor violating processes are not generic features of low scale seesaw models. The $S_4$ flavor symmetry and the different auxiliary cylic symmetries introduced in our model, allows to get defined predictions for the branching ratios for the lepton flavor violating decays $\mu \rightarrow e\gamma $, $\tau\rightarrow \mu \gamma $ and $\tau \rightarrow e\gamma $. Depending on the discrete symmetries assignments one can have, for instance sizeable $\tau\to e\gamma$ decay, but strongly suppressed $\mu\to e\gamma$ and $\tau\to\mu\gamma$ processes as shown in the $A_4$ flavor model of Ref. \cite{CarcamoHernandez:2019pmy}.

\begin{figure}[H]
	\begin{subfigure}{.5\linewidth}
		\centering
		\captionsetup{width=0.9\textwidth}
		\includegraphics[scale=.6]{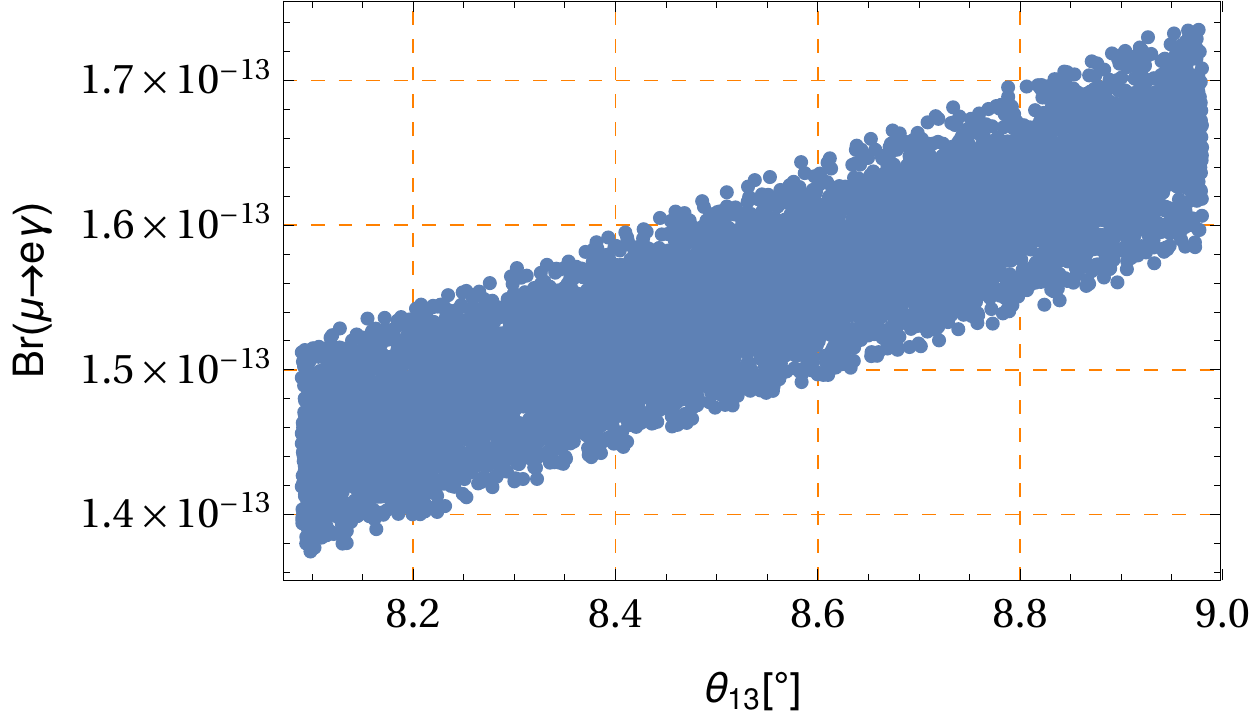}
		\caption{Correlation between $Br(\mu\to e\gamma)$ and the reactor mixing angle $\theta_{13}$.}
		\label{fig:sub1}
	\end{subfigure}
	\begin{subfigure}{.5\linewidth}
		\centering
		\captionsetup{width=0.9\textwidth}
		\includegraphics[scale=.6]{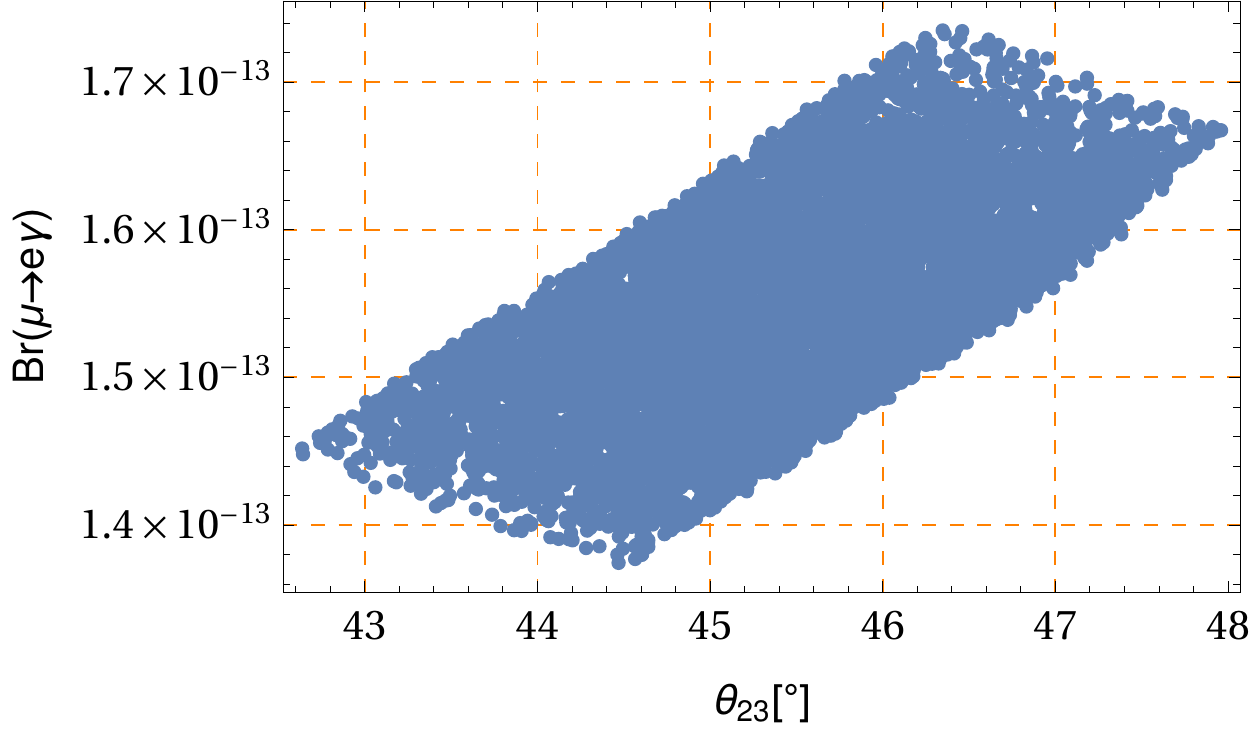}
		\caption{Correlation between $Br(\mu\to e\gamma)$ and the atmospheric mixing angle $\theta_{23}$.}
		\label{fig:sub2}
	\end{subfigure}\\
	\begin{subfigure}{.5\linewidth}
		\centering
		\captionsetup{width=0.9\textwidth}
		\includegraphics[scale=.6]{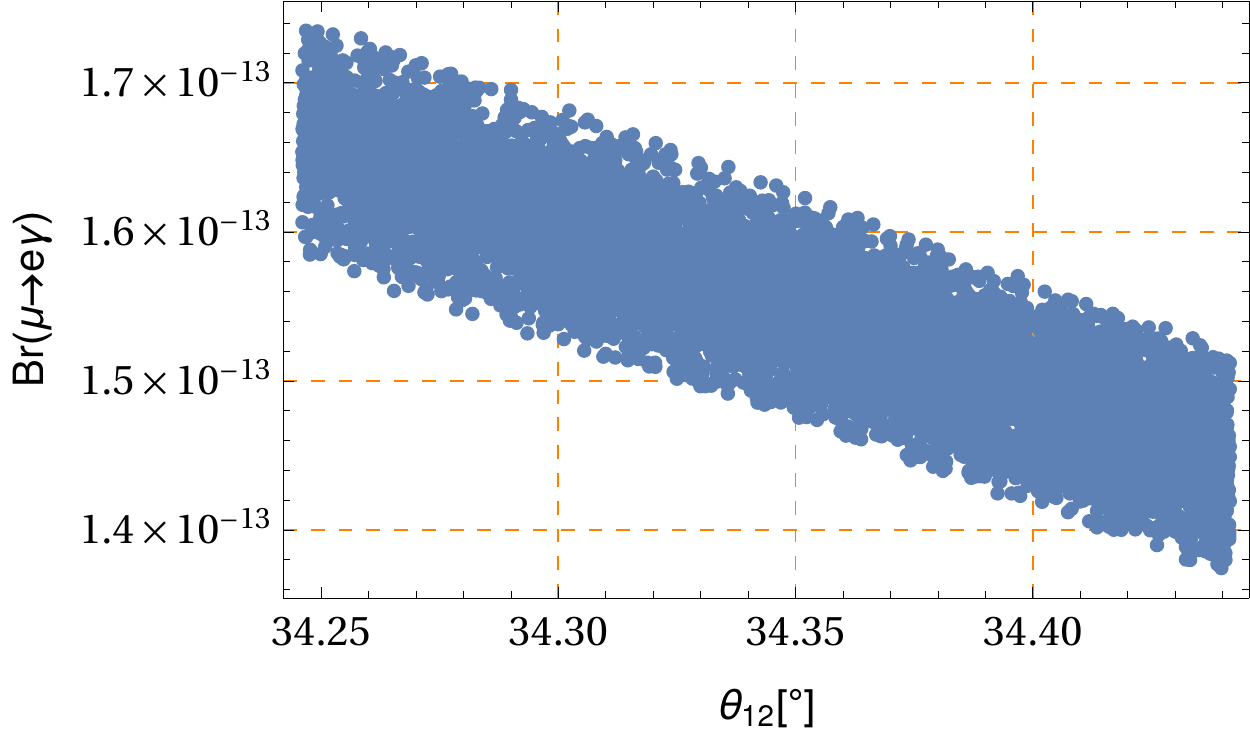}
		\caption{Correlation between $Br(\mu\to e\gamma)$ and solar mixing angle $\theta_{12}$.}
		\label{fig:sub3}
	\end{subfigure}
	\begin{subfigure}{.5\linewidth}
		\centering
		\captionsetup{width=0.9\textwidth}
		\includegraphics[scale=.6]{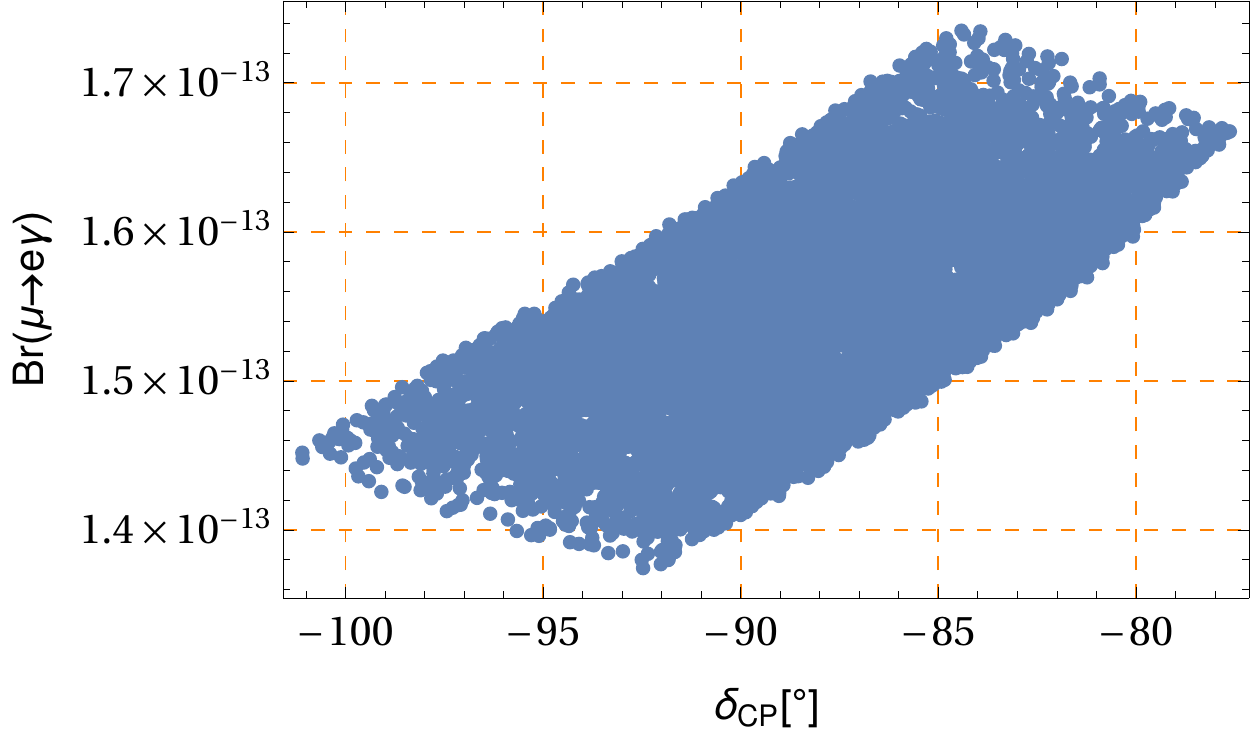}
		\caption{Correlation between $Br(\mu\to e\gamma)$ and the leptonic Dirac CP violating phase $\delta_{CP}$.}
		\label{fig:sub3}
	\end{subfigure}
	\caption{Correlations between $Br(\mu\to e\gamma)$ and the different lepton sector observables.} 
	\label{fig:correlationBrwiththetaijl}
\end{figure}

Figure \ref{fig:correlationBrwiththetaijl} shows the correlations of the Branching ratio for the $\mu\to e\gamma$ decay with the leptonic mixing angles as well as with the leptonic Dirac CP violating phase. To obtain these Figures, the lepton sector parameters were randomly generated in a range of values where the neutrino mass squared splittings, leptonic mixing angles and leptonic Dirac CP violating phase are inside the $3\sigma$ experimentally allowed range. The plots in Figure \ref{fig:correlationBrwiththetaijl} show that the Branching ratio for the $\mu\to e\gamma$ decay increases when the reactor $\theta_{13}$ and atmospheric $\theta_{23}$ mixing angles as well as the leptonic Dirac CP violating phase $\delta_{CP}$ take larger values. On the other hand, the Branching ratio for the $\mu\to e\gamma$ decay decreases as the solar mixing angle $\theta_{12}$ is increased.

\newpage

\section{Conclusions}

We have proposed a minimal predictive inverse seesaw model based on two
right-handed neutrinos and two additional singlets, which yields the same
low energy neutrino mass matrix as in the Littlest Seesaw (LS) (type I)
model. The model is called the Littlest Inverse Seesaw (LIS) model and
yields the mass matrix structures as shown in the Introduction.

In order to implement the LIS model, we have used an $S_{4}$ family
symmetry, supplemented by the $U\left( 1\right) \times Z_{3}\times
Z_{6}\times U\left( 1\right) _{R}$ group. The charged lepton mass hierarchy
is produced by the spontaneous breaking of the $S_4\times Z_{3}\times Z_{6}$
discrete group at very high energies. The nature of the inverse seesaw
mechanism is guaranteed by renormalizable and non-renormalizable mass terms
involving gauge singlet right handed Majorana neutrinos. These terms are
generated after the spontaneous breaking of the $U(1)$ global symmetry at
the TeV scale.

The resulting LIS model proposed here is the first low scale seesaw model
which incorporates the successful predictions of the LS model, including the
prediction of a normal neutrino mass ordering, all arising from only two
effective free parameters. However there is one crucial phenomenological
difference between the LS and the LIS models: the LIS model allows charged
lepton flavour violating (CLFV) processes within the reach of future
experimental sensitivity. In addition, we have studied the implications of our model in the lepton conversion in nuclei. We have found that in order that our model’s predictions for the $CR\left(\mu Ti\rightarrow eTi\right)$ and $CR\left(\mu Al\rightarrow eAl\right)$ effective parameters be lower than the expected sensitivities of the next generation of experiments that will use Titanium and Aluminum as targets, the effective neutrino Yukawa coupling $x_{2}^{(\nu)}$ has to be lower than about $0.2$ and $0.4$, respectively, for sterile neutrino masses larger than around $300$ GeV.

In summary, the LIS model predicts branching ratios for the charged lepton
flavour violating processes: $\mu \rightarrow e\gamma $, $\tau\rightarrow
\mu \gamma $ and $\tau \rightarrow e\gamma $ in the ranges $8\times
10^{-14}\lesssim Br\left( \mu \rightarrow e\gamma \right) \lesssim 1.8\times
10^{-13}$, $2\times 10^{-13}\lesssim Br\left( \tau \rightarrow \mu \gamma
\right) \lesssim 1.6\times 10^{-12}$ and $2\times 10^{-14}\lesssim Br\left(
\tau \rightarrow e\gamma \right)\lesssim 1.8\times 10^{-13}$, which will all
present a target for the forthcoming CLFV experiments.

\label{conclusions}

\section*{Acknowledgments}

This research has received funding from Fondecyt (Chile), Grants
No.~1170803, CONICYT PIA/Basal FB0821. A.E.C.H thanks University of
Southampton and Institute of Experimental and Applied Physics of the Czech Technical University in Praga, where part of this work was done. SFK
acknowledges the STFC Consolidated Grant ST/L000296/1 and the European
Union's Horizon 2020 Research and Innovation programme under Marie Sk\l %
{}odowska-Curie grant agreements Elusives ITN No.\ 674896 and InvisiblesPlus
RISE No.\ 690575. \appendix

\section{$S_4$ Symmetry}

\label{S4}The $S_{4}$ is the smallest non abelian group having doublet and
singlet irreducible representations. $S_{4}$ is the group of permutations of
four objects, which includes five irreducible representations, i.e., $%
\mathbf{1,1^{\prime },2,3,3^{\prime }}$ fulfulling the following tensor
product rules \cite{Ishimori:2010au} 
\begin{align}
& \mathbf{3}\otimes \mathbf{3}=\mathbf{1}\oplus \mathbf{2}\oplus \mathbf{3}%
\oplus \mathbf{3^{\prime }},\qquad \mathbf{3^{\prime }}\otimes \mathbf{%
3^{\prime }}=\mathbf{1}\oplus \mathbf{2}\oplus \mathbf{3}\oplus \mathbf{%
3^{\prime }},\qquad \mathbf{3}\otimes \mathbf{3^{\prime }}=\mathbf{1^{\prime
}}\oplus \mathbf{2}\oplus \mathbf{3}\oplus \mathbf{3^{\prime }}, \\
& \mathbf{2}\otimes \mathbf{2}=\mathbf{1}\oplus \mathbf{1^{\prime }}\oplus 
\mathbf{2},\qquad \mathbf{2}\otimes \mathbf{3}=\mathbf{3}\oplus \mathbf{%
3^{\prime }},\qquad \mathbf{2}\otimes \mathbf{3^{\prime }}=\mathbf{3^{\prime
}}\oplus \mathbf{3}, \\
& \mathbf{3}\otimes \mathbf{1^{\prime }}=\mathbf{3^{\prime }},\qquad \mathbf{%
3^{\prime }}\otimes \mathbf{1^{\prime }}=\mathbf{3},\qquad \mathbf{2}\otimes 
\mathbf{1^{\prime }}=\mathbf{2}.
\end{align}%
Explicitly, the basis used in this paper corresponds to Ref. \cite%
{Ishimori:2010au} and results in 
\begin{equation}
(\mathbf{A})_{\mathbf{3}}\times (\mathbf{B})_{\mathbf{3}}=(\mathbf{A}\cdot 
\mathbf{B})_{\mathbf{1}}+\left( 
\begin{array}{c}
\mathbf{A}\cdot \Sigma \cdot \mathbf{B} \\ 
\mathbf{A}\cdot \Sigma ^{\ast }\cdot \mathbf{B}%
\end{array}%
\right) _{\mathbf{2}}+\left( 
\begin{array}{c}
\{A_{y}B_{z}\} \\ 
\{A_{z}B_{x}\} \\ 
\{A_{x}B_{y}\}%
\end{array}%
\right) _{\mathbf{3}}+\left( 
\begin{array}{c}
\left[ A_{y}B_{z}\right] \\ 
\left[ A_{z}B_{x}\right] \\ 
\left[ A_{x}B_{y}\right]%
\end{array}%
\right) _{\mathbf{3^{\prime }}},
\end{equation}%
\begin{equation}
(\mathbf{A})_{\mathbf{3^{\prime }}}\times (\mathbf{B})_{\mathbf{3^{\prime }}%
}=(\mathbf{A}\cdot \mathbf{B})_{\mathbf{1}}+\left( 
\begin{array}{c}
\mathbf{A}\cdot \Sigma \cdot \mathbf{B} \\ 
\mathbf{A}\cdot \Sigma ^{\ast }\cdot \mathbf{B}%
\end{array}%
\right) _{\mathbf{2}}+\left( 
\begin{array}{c}
\{A_{y}B_{z}\} \\ 
\{A_{z}B_{x}\} \\ 
\{A_{x}B_{y}\}%
\end{array}%
\right) _{\mathbf{3}}+\left( 
\begin{array}{c}
\left[ A_{y}B_{z}\right] \\ 
\left[ A_{z}B_{x}\right] \\ 
\left[ A_{x}B_{y}\right]%
\end{array}%
\right) _{\mathbf{3^{\prime }}},
\end{equation}%
\begin{equation}
(\mathbf{A})_{\mathbf{3}}\times (\mathbf{B})_{\mathbf{3^{\prime }}}=(\mathbf{%
A}\cdot \mathbf{B})_{\mathbf{1^{\prime }}}+\left( 
\begin{array}{c}
\mathbf{A}\cdot \Sigma \cdot \mathbf{B} \\ 
-\mathbf{A}\cdot \Sigma ^{\ast }\cdot \mathbf{B}%
\end{array}%
\right) _{\mathbf{2}}+\left( 
\begin{array}{c}
\{A_{y}B_{z}\} \\ 
\{A_{z}B_{x}\} \\ 
\{A_{x}B_{y}\}%
\end{array}%
\right) _{\mathbf{3^{\prime }}}+\left( 
\begin{array}{c}
\left[ A_{y}B_{z}\right] \\ 
\left[ A_{z}B_{x}\right] \\ 
\left[ A_{x}B_{y}\right]%
\end{array}%
\right) _{\mathbf{3}},
\end{equation}%
\begin{equation}
(\mathbf{A})_{\mathbf{2}}\times (\mathbf{B})_{\mathbf{2}}=\{A_{x}B_{y}\}_{%
\mathbf{1}}+\left[ A_{x}B_{y}\right] _{\mathbf{1^{\prime }}}+\left( 
\begin{array}{c}
A_{y}B_{y} \\ 
A_{x}B_{x}%
\end{array}%
\right) _{\mathbf{2}},
\end{equation}%
\begin{equation}
\left( 
\begin{array}{c}
A_{x} \\ 
A_{y}%
\end{array}%
\right) _{\mathbf{2}}\times \left( 
\begin{array}{c}
B_{x} \\ 
B_{y} \\ 
B_{z}%
\end{array}%
\right) _{\mathbf{3}}=\left( 
\begin{array}{c}
(A_{x}+A_{y})B_{x} \\ 
(\omega ^{2}A_{x}+\omega A_{y})B_{y} \\ 
(\omega A_{x}+\omega ^{2}A_{y})B_{z}%
\end{array}%
\right) _{\mathbf{3}}+\left( 
\begin{array}{c}
(A_{x}-A_{y})B_{x} \\ 
(\omega ^{2}A_{x}-\omega A_{y})B_{y} \\ 
(\omega A_{x}-\omega ^{2}A_{y})B_{z}%
\end{array}%
\right) _{\mathbf{3}^{\prime }},
\end{equation}%
\begin{equation}
\left( 
\begin{array}{c}
A_{x} \\ 
A_{y}%
\end{array}%
\right) _{\mathbf{2}}\times \left( 
\begin{array}{c}
B_{x} \\ 
B_{y} \\ 
B_{z}%
\end{array}%
\right) _{\mathbf{3}^{\prime }}=\left( 
\begin{array}{c}
(A_{x}+A_{y})B_{x} \\ 
(\omega ^{2}A_{x}+\omega A_{y})B_{y} \\ 
(\omega A_{x}+\omega ^{2}A_{y})B_{z}%
\end{array}%
\right) _{\mathbf{3}^{\prime }}+\left( 
\begin{array}{c}
(A_{x}-A_{y})B_{x} \\ 
(\omega ^{2}A_{x}-\omega A_{y})B_{y} \\ 
(\omega A_{x}-\omega ^{2}A_{y})B_{z}%
\end{array}%
\right) _{\mathbf{3}},
\end{equation}%
with 
\begin{align}
\mathbf{A}\cdot \mathbf{B}& =A_{x}B_{x}+A_{y}B_{y}+A_{z}B_{z},  \notag \\
\{A_{x}B_{y}\}& =A_{x}B_{y}+A_{y}B_{x},  \notag \\
\left[ A_{x}B_{y}\right] & =A_{x}B_{y}-A_{y}B_{x},  \notag \\
\mathbf{A}\cdot \Sigma \cdot \mathbf{B}& =A_{x}B_{x}+\omega
A_{y}B_{y}+\omega ^{2}A_{z}B_{z},  \notag \\
\mathbf{A}\cdot \Sigma ^{\ast }\cdot \mathbf{B}& =A_{x}B_{x}+\omega
^{2}A_{y}B_{y}+\omega A_{z}B_{z},
\end{align}%
where $\omega =e^{2\pi i/3}$ is a complex square root of unity.

\section{The $S_{4}$ flavored superpotential.}

\label{WS4} In order to obtain the VEV configuration for the $S_{4}$ doublet
and triplet scalars shown in Eq. (\ref{VEV}), we consider the following $%
S_{4}\times U\left( 1\right) \times Z_{3}\times Z_{6}$ invariant
superpotential: 
\begin{eqnarray}
W &=&\kappa _{1}\left( \eta \eta \right) _{\mathbf{1}}X_{1}+\kappa
_{2}\left( \phi \phi \right) _{\mathbf{1}}X_{2}+\kappa _{3}\left( \sigma
_{\mu }\sigma _{\tau }\right) _{\mathbf{1}}X_{3}+\kappa _{4}\left( \sigma
_{e}\sigma _{\tau }\right) _{\mathbf{1}}X_{4}+\kappa _{5}\left( \sigma
_{e}\sigma _{\mu }\right) _{\mathbf{1}}X_{5}+\frac{\kappa _{6}}{\Lambda }%
\left[ \left( \eta \eta \right) _{\mathbf{2}}\phi \right] _{\mathbf{1}%
^{\prime }}X_{6}  \notag \\
&&+\frac{\kappa _{7}}{\Lambda }\left[ \left( \chi \chi \right) _{\mathbf{3}%
}\sigma _{\mu }\right] _{\mathbf{1}}X_{7}+\frac{\kappa _{8}}{\Lambda }\left[
\left( \chi \chi \right) _{\mathbf{3}}\sigma _{\tau }\right] _{\mathbf{1}%
}X_{8}+\frac{\kappa _{9}}{\Lambda }\left[ \left( \varphi \varphi \right) _{%
\mathbf{2}}\varphi \right] _{\mathbf{1}^{\prime }}X_{9}+\frac{\kappa _{10}}{%
\Lambda }\left[ \left( \sigma _{\mu }\sigma _{\mu }\right) _{\mathbf{2}%
}\varphi \right] _{\mathbf{1}}X_{10}  \notag \\
&&+\frac{\kappa _{11}}{\Lambda }\left[ \left( \xi \xi \right) _{\mathbf{%
\mathbf{2}}}\varphi \right] _{\mathbf{1}^{\prime }}X_{11}+\kappa _{12}\left(
\sigma _{\mu }\sigma _{\mu }\right) _{\mathbf{3}}\Phi +\kappa _{13}\left(
\sigma _{\tau }\sigma _{\tau }\right) _{\mathbf{3}}\Delta +\kappa
_{14}\left( \sigma _{e}\sigma _{e}\right) _{\mathbf{3}}\Theta +\frac{\kappa
_{15}}{\Lambda }\left[ \left( \chi \chi \right) _{\mathbf{\mathbf{2}}}\sigma
_{e}\right] _{\mathbf{3}^{\prime }}\Xi  \notag \\
&=&2\kappa _{1}\eta _{1}\eta _{2}X_{1}+2\kappa _{2}\phi _{1}\phi
_{2}X_{2}+\kappa _{3}\left( \sigma _{1\mu }\sigma _{1\tau }+\sigma _{2\mu
}\sigma _{2\tau }+\sigma _{3\mu }\sigma _{3\tau }\right) X_{3}+\kappa
_{4}\left( \sigma _{1e}\sigma _{1\tau }+\sigma _{2e}\sigma _{2\tau }+\sigma
_{3e}\sigma _{3\tau }\right) X_{4}  \notag \\
&&+\kappa _{5}\left( \sigma _{1e}\sigma _{1\mu }+\sigma _{2e}\sigma _{2\mu
}+\sigma _{3e}\sigma _{3\mu }\right) X_{5}+\frac{\kappa _{6}}{\Lambda }%
\left( \eta _{2}^{2}\phi _{2}-\eta _{1}^{2}\phi _{1}\right) X_{6}+\frac{%
\kappa _{7}}{\Lambda }\left( \sigma _{1\mu }\chi _{2}\chi _{3}+\sigma _{2\mu
}\chi _{1}\chi _{3}+\sigma _{3\mu }\chi _{1}\chi _{2}\right) X_{7}  \notag \\
&&+\frac{\kappa _{8}}{\Lambda }\left( \sigma _{1\tau }\chi _{2}\chi
_{3}+\sigma _{2\tau }\chi _{1}\chi _{3}+\sigma _{3\tau }\chi _{1}\chi
_{2}\right) X_{8}+\frac{\kappa _{9}}{\Lambda }\left( \varphi
_{2}^{3}-\varphi _{1}^{3}\right) X_{9}  \notag \\
&&+\frac{\kappa _{10}}{\Lambda }\left[ \varphi _{2}\left( \sigma _{1\mu
}^{2}+\omega \sigma _{2\mu }^{2}+\omega ^{2}\sigma _{3\mu }^{2}\right)
+\varphi _{1}\left( \sigma _{1\mu }^{2}+\omega ^{2}\sigma _{2\mu
}^{2}+\omega \sigma _{3\mu }^{2}\right) \right] X_{10}  \notag \\
&&+\frac{\kappa _{11}}{\Lambda }\left[ \varphi _{2}\left( \xi
_{1}^{2}+\omega \xi _{2}^{2}+\omega ^{2}\xi _{3}^{2}\right) -\varphi
_{1}\left( \xi _{1}^{2}+\omega ^{2}\xi _{2}^{2}+\omega \xi _{3}^{2}\right) %
\right] X_{11}  \notag \\
&&+2\kappa _{12}\left( \sigma _{2\mu }\sigma _{3\mu }\Phi _{1}+\sigma _{1\mu
}\sigma _{3\mu }\Phi _{2}+\sigma _{1\mu }\sigma _{2\mu }\Phi _{3}\right)
+2\kappa _{13}\left( \sigma _{2\tau }\sigma _{3\tau }\Delta _{1}+\sigma
_{1\tau }\sigma _{3\tau }\Delta _{2}+\sigma _{1\tau }\sigma _{2\tau }\Delta
_{3}\right)  \notag \\
&&+2\kappa _{14}\left( \sigma _{2e}\sigma _{3e}\Theta _{1}+\sigma
_{1e}\sigma _{3e}\Theta _{2}+\sigma _{1e}\sigma _{2e}\Theta _{3}\right) 
\notag \\
&&+\frac{\kappa _{15}}{\Lambda }\left[ \left( \chi _{1}^{2}+\omega \chi
_{2}^{2}+\omega ^{2}\chi _{3}^{2}\right) -\left( \chi _{1}^{2}+\omega
^{2}\chi _{2}^{2}+\omega \chi _{3}^{2}\right) \right] \sigma _{1e}\Xi _{1} 
\notag \\
&&+\frac{\kappa _{15}}{\Lambda }\left[ \omega ^{2}\left( \chi
_{1}^{2}+\omega \chi _{2}^{2}+\omega ^{2}\chi _{3}^{2}\right) -\omega \left(
\chi _{1}^{2}+\omega ^{2}\chi _{2}^{2}+\omega \chi _{3}^{2}\right) \right]
\sigma _{2e}\Xi _{2}  \notag \\
&&+\frac{\kappa _{15}}{\Lambda }\left[ \omega \left( \chi _{1}^{2}+\omega
\chi _{2}^{2}+\omega ^{2}\chi _{3}^{2}\right) -\omega ^{2}\left( \chi
_{1}^{2}+\omega ^{2}\chi _{2}^{2}+\omega \chi _{3}^{2}\right) \right] \sigma
_{3e}\Xi _{3}  \label{Superpotential}
\end{eqnarray}
Notice that there are two scales for the VEVs of the gauge singlet scalar fields of our model, i.e., the TeV scale and the large scale $\approx\lambda\Lambda$, with $\lambda=0.225$. The singlet scalar fields having TeV scale VEVs are charged under the global $U(1)$ symmetry, whereas the remaining scalar singlets are neutral under this symmetry and do acquire VEVs at the large scale. Because of this reason higher order terms in the superpotential will not affect the stability of the VEVs.

From the superpotential given above, we find the following potential minimum
conditions: 
\begin{eqnarray}
v_{\eta _{1}}v_{\eta _{2}} &=&0,\hspace{1cm}\hspace{1cm}v_{\phi _{1}}v_{\phi
_{2}}=0,  \label{W1} \\
v_{\sigma _{1\mu }}v_{\chi _{2}}v_{\chi _{3}}+v_{\sigma _{2\mu }}v_{\chi
_{1}}v_{\chi _{3}}+v_{\sigma _{3\mu }}v_{\chi _{1}}v_{\chi _{2}} &=&0,%
\hspace{1cm}\hspace{1cm}v_{\eta _{2}}^{2}v_{\phi _{2}}-v_{\eta
_{1}}^{2}v_{\phi _{1}}=0,  \label{W2} \\
v_{\sigma _{1\tau }}v_{\chi _{2}}v_{\chi _{3}}+v_{\sigma _{2\tau }}v_{\chi
_{1}}v_{\chi _{3}}+v_{\sigma _{3\tau }}v_{\chi _{1}}v_{\chi _{2}} &=&0,
\label{W3} \\
v_{\sigma _{1\mu }}v_{\sigma _{1\tau }}+v_{\sigma _{2\mu }}v_{\sigma _{2\tau
}}+v_{\sigma _{3\mu }}v_{\sigma _{3\tau }} &=&0,\hspace{1cm}\hspace{1cm}%
v_{\varphi _{2}}^{3}-v_{\varphi _{1}}^{3}=0,  \label{W4} \\
v_{\sigma _{1e}}v_{\sigma _{1\tau }}+v_{\sigma _{2e}}v_{\sigma _{2\tau
}}+v_{\sigma _{3e}}v_{\sigma _{3\tau }} &=&0,  \label{W4a} \\
v_{\sigma _{1\mu }}v_{\sigma _{1e}}+v_{\sigma _{2\mu }}v_{\sigma
_{2e}}+v_{\sigma _{3\mu }}v_{\sigma _{3e}} &=&0,  \label{W4b} \\
v_{\varphi _{2}}\left( v_{\sigma _{1\mu }}^{2}+\omega v_{\sigma _{2\mu
}}^{2}+\omega ^{2}v_{\sigma _{3\mu }}^{2}\right) +v_{\varphi _{1}}\left(
v_{\sigma _{1\mu }}^{2}+\omega ^{2}v_{\sigma _{2\mu }}^{2}+\omega v_{\sigma
_{3\mu }}^{2}\right) &=&0,  \label{W5} \\
v_{\varphi _{2}}\left( v_{\xi _{1}}^{2}+\omega v_{\xi _{2}}^{2}+\omega
^{2}v_{\xi _{3}}^{2}\right) -v_{\varphi _{1}}\left( v_{\xi _{1}}^{2}+\omega
^{2}v_{\xi _{2}}^{2}+\omega v_{\xi _{3}}^{2}\right) &=&0,  \label{W6} \\
v_{\chi _{1}}v_{\zeta _{1}}+v_{\chi _{2}}v_{\zeta _{2}}+v_{\chi
_{3}}v_{\zeta _{3}} &=&0,\hspace{1cm}v_{\xi _{1}}v_{\zeta _{1}}+v_{\xi
_{2}}v_{\zeta _{2}}+v_{\xi _{3}}v_{\zeta _{3}}=0,  \label{W7} \\
v_{\sigma _{2\mu }}v_{\sigma _{1\mu }} &=&0,\hspace{1cm}\hspace{1cm}%
v_{\sigma _{3\mu }}v_{\sigma _{2\mu }}=0,  \label{W8} \\
v_{\sigma _{1\mu }}v_{\sigma _{3\mu }} &=&0,\hspace{1cm}\hspace{1cm}%
v_{\sigma _{2\tau }}v_{\sigma _{1\tau }}=0,  \label{W9} \\
v_{\sigma _{3\tau }}v_{\sigma _{2\tau }} &=&0,\hspace{1cm}\hspace{1cm}%
v_{\sigma _{1\tau }}v_{\sigma _{3\tau }}=0,  \label{W10} \\
\left[ \left( v_{\chi _{1}}^{2}+\omega v_{\chi _{2}}^{2}+\omega ^{2}v_{\chi
_{3}}^{2}\right) -\left( v_{\chi _{1}}^{2}+\omega ^{2}v_{\chi
_{2}}^{2}+\omega v_{\chi _{3}}^{2}\right) \right] v_{\sigma _{1e}} &=&0,
\label{W11} \\
\left[ \omega ^{2}\left( v_{\chi _{1}}^{2}+\omega v_{\chi _{2}}^{2}+\omega
^{2}v_{\chi _{3}}^{2}\right) -\omega \left( v_{\chi _{1}}^{2}+\omega
^{2}v_{\chi _{2}}^{2}+\omega v_{\chi _{3}}^{2}\right) \right] v_{\sigma
_{2e}} &=&0,  \label{W12} \\
\left[ \omega \left( v_{\chi _{1}}^{2}+\omega v_{\chi _{2}}^{2}+\omega
^{2}v_{\chi _{3}}^{2}\right) -\omega ^{2}\left( v_{\chi _{1}}^{2}+\omega
^{2}v_{\chi _{2}}^{2}+\omega v_{\chi _{3}}^{2}\right) \right] v_{\sigma
_{3e}} &=&0.  \label{W13}
\end{eqnarray}

Combining Eqs (\ref{W1}) and (\ref{W2}) we find:
\begin{equation}
v_{\eta _{2}}=v_{\phi _{1}}=0,\hspace{1cm}v_{\eta _{1}}\neq 0,\hspace{1cm}%
v_{\phi _{2}}\neq 0,\hspace{1cm}\mbox{or}\hspace{1cm}v_{\eta _{1}}=v_{\phi
_{2}}=0,\hspace{1cm}v_{\eta _{2}}\neq 0,\hspace{1cm}v_{\phi _{1}}\neq 0
\label{S1}
\end{equation}

Furthermore, from Eq. (\ref{W4}), we get:
\begin{equation}
v_{\varphi _{2}}=v_{\varphi _{1}},\hspace{1cm}v_{\varphi _{2}}=\omega ^{\pm
1}v_{\varphi _{1}},\hspace{1cm}\mbox{or}\hspace{1cm}v_{\varphi _{1}}=\omega
^{\pm 1}v_{\varphi _{2}}  \label{S2}
\end{equation}

We proceed to choose the solution:
\begin{equation}
\left\langle \varphi \right\rangle =v_{\varphi }\left( 1,\omega \right) ,%
\hspace{1cm}\left\langle \phi \right\rangle =v_{\phi }\left( 0,1\right) ,%
\hspace{1cm}\left\langle \eta \right\rangle =v_{\eta }\left( 1,0\right) ,
\label{S3}
\end{equation}

Then, using the above given VEV configuration for $\varphi $, Eqs. (\ref{W5}%
) and (\ref{W6}) take the form:

\begin{equation}
v_{\sigma _{1\mu }}^{2}+v_{\sigma _{3\mu }}^{2}=0,\hspace{1cm}\hspace{1cm}%
v_{\xi _{1}}^{2}-v_{\xi _{3}}^{2}=0,  \label{S5}
\end{equation}

Restricting to real solutions for the components of the VEV patterns for the 
$S_{4}$ scalar triplets, from Eqs. (\ref{S5}), (\ref{W8}) and (\ref{W9}), we
find:
\begin{equation}
\left\langle \sigma _{\mu }\right\rangle =v_{\sigma _{\mu }}\left(
0,1,0\right) ,  \label{S6}
\end{equation}

Thus, replacing Eq. (\ref{S6}) in Eqs. (\ref{W4a}) and (\ref{W4b}) yield the
following VEV pattern for the $S_{4}$ scalar triplet $\sigma _{e}$:

\begin{equation}
\left\langle \sigma _{e}\right\rangle =v_{\sigma _{e}}\left( 1,0,0\right) .
\label{S6c}
\end{equation}

Furthermore, from Eq. (\ref{S5}), we find:
\begin{equation}
\left\langle \xi \right\rangle =v_{\xi }\left( \pm 1,r,\pm 1\right) ,\hspace{%
1cm}\mbox{or}\hspace{1cm}\left\langle \xi \right\rangle =v_{\xi }\left( \pm
1,r,\mp 1\right) .  \label{S6a}
\end{equation}

We choose the following solution:
\begin{equation}
\left\langle \xi \right\rangle =v_{\xi }\left( 1,3,1\right) .  \label{S6b}
\end{equation}

Replacing the VEV pattern of the $S_{4}$ triplet $\sigma _{\mu }$ given by
Eq. (\ref{S6}) in Eqs. (\ref{W4}), (\ref{W9}) and (\ref{W10}), we get:
\begin{equation}
\left\langle \sigma _{\tau }\right\rangle =v_{\sigma _{\tau }}\left(
0,0,1\right) .  \label{S7}
\end{equation}

Furthermore, combining Eqs. (\ref{W2}), (\ref{W3}), (\ref{S6}) and (\ref{S7}%
) yield the following relations: 
\begin{equation}
v_{\chi _{1}}v_{\chi _{3}}=0,\hspace{1cm}\hspace{1cm}v_{\chi _{1}}v_{\chi
_{2}}=0,  \label{S8}
\end{equation}

which implies one of the following solutions:
\begin{equation}
v_{\chi _{1}}=0,\hspace{1cm}v_{\chi _{2}}\neq 0,\hspace{1cm}v_{\chi
_{3}}\neq 0,\hspace{1cm}\mbox{or}\hspace{1cm}v_{\chi _{1}}=v_{\chi
_{2}}=v_{\chi _{3}}=0.  \label{S9}
\end{equation}

We choose the nontrivial solution:
\begin{equation}
v_{\chi _{1}}=0,\hspace{1cm}v_{\chi _{2}}\neq 0,\hspace{1cm}v_{\chi
_{3}}\neq 0.  \label{S10}
\end{equation}

Besides that, from Eqs. (\ref{S6c}) and (\ref{W11}), we find:
\begin{equation}
v_{\chi _{2}}^{2}-v_{\chi _{3}}^{2}=0.  \label{S11}
\end{equation}

Consequently, the $S_{4}$ triplet $\chi $ has the following VEV configuration:
\begin{equation}
\left\langle \chi \right\rangle =v_{\chi }\left( 0,1,1\right) .  \label{S12}
\end{equation}

Thus, the potential minimum conditions yield the following VEV patterns for
the $S_{4}$ doublets and triplet scalars of our model:
\begin{eqnarray}
\left\langle \varphi \right\rangle &=&v_{\varphi }\left( 1,\omega \right) ,%
\hspace{1cm}\left\langle \phi \right\rangle =v_{\phi }\left( 0,1\right) ,%
\hspace{1cm}\left\langle \eta \right\rangle =v_{\eta }\left( 1,0\right) ,%
\hspace{1cm}\left\langle \chi \right\rangle =v_{\chi }\left( 0,1,1\right) , 
\notag \\
\left\langle \xi \right\rangle &=&v_{\xi }\left( 1,3,1\right) ,\hspace{1cm}%
\left\langle \sigma _{\mu }\right\rangle =v_{\sigma _{\mu }}\left(
0,1,0\right) ,\hspace{1cm}\left\langle \sigma _{\tau }\right\rangle
=v_{\sigma _{\tau }}\left( 0,0,1\right) ,\hspace{1cm}\left\langle \sigma
_{e}\right\rangle =v_{\sigma _{e}}\left( 1,0,0\right) .  \label{VEVconfig}
\end{eqnarray}.


\begin{thebibliography}{9}
\bibitem{Minkowski:1977sc} 
  P.~Minkowski,
  Phys.\ Lett.\  {\bf 67B}, 421 (1977).
  doi:10.1016/0370-2693(77)90435-X



\bibitem{Yanagida:1979as} 
  T.~Yanagida,
  Conf.\ Proc.\ C {\bf 7902131}, 95 (1979).



\bibitem{GellMann:1980vs} 
  M.~Gell-Mann, P.~Ramond and R.~Slansky,
  Conf.\ Proc.\ C {\bf 790927}, 315 (1979)
  [arXiv:1306.4669 [hep-th]].



\bibitem{Glashow:1979nm} 
  S.~L.~Glashow,
  NATO Sci.\ Ser.\ B {\bf 61}, 687 (1980).



\bibitem{Mohapatra:1979ia} 
  R.~N.~Mohapatra and G.~Senjanovic,
  Phys.\ Rev.\ Lett.\  {\bf 44}, 912 (1980).
  doi:10.1103/PhysRevLett.44.912



\bibitem{King:1999mb} 
  S.~F.~King,
  Nucl.\ Phys.\ B {\bf 576}, 85 (2000)
  doi:10.1016/S0550-3213(00)00109-7
  [hep-ph/9912492].



\bibitem{King:2002nf} 
  S.~F.~King,
  JHEP {\bf 0209}, 011 (2002)
  doi:10.1088/1126-6708/2002/09/011
  [hep-ph/0204360].



\bibitem{Frampton:2002qc} 
  P.~H.~Frampton, S.~L.~Glashow and T.~Yanagida,
  Phys.\ Lett.\ B {\bf 548}, 119 (2002)
  doi:10.1016/S0370-2693(02)02853-8
  [hep-ph/0208157].



\bibitem{Fukugita:1986hr} 
  M.~Fukugita and T.~Yanagida,
  Phys.\ Lett.\ B {\bf 174}, 45 (1986).
  doi:10.1016/0370-2693(86)91126-3



\bibitem{Guo:2003cc} 
  W.~l.~Guo and Z.~z.~Xing,
  Phys.\ Lett.\ B {\bf 583}, 163 (2004)
  doi:10.1016/j.physletb.2003.12.043
  [hep-ph/0310326].



\bibitem{Ibarra:2003up} 
  A.~Ibarra and G.~G.~Ross,
  Phys.\ Lett.\ B {\bf 591}, 285 (2004)
  doi:10.1016/j.physletb.2004.04.037
  [hep-ph/0312138].



\bibitem{Mei:2003gn} 
  J.~w.~Mei and Z.~z.~Xing,
  Phys.\ Rev.\ D {\bf 69}, 073003 (2004)
  doi:10.1103/PhysRevD.69.073003
  [hep-ph/0312167].



\bibitem{Guo:2006qa} 
  W.~l.~Guo, Z.~z.~Xing and S.~Zhou,
  Int.\ J.\ Mod.\ Phys.\ E {\bf 16}, 1 (2007)
  doi:10.1142/S0218301307004898
  [hep-ph/0612033].



\bibitem{Antusch:2011nz} 
  S.~Antusch, P.~Di Bari, D.~A.~Jones and S.~F.~King,
  Phys.\ Rev.\ D {\bf 86}, 023516 (2012)
  doi:10.1103/PhysRevD.86.023516
  [arXiv:1107.6002 [hep-ph]].



\bibitem{Harigaya:2012bw} 
  K.~Harigaya, M.~Ibe and T.~T.~Yanagida,
  Phys.\ Rev.\ D {\bf 86}, 013002 (2012)
  doi:10.1103/PhysRevD.86.013002
  [arXiv:1205.2198 [hep-ph]].



\bibitem{Zhang:2015tea} 
  J.~Zhang and S.~Zhou,
  JHEP {\bf 1509}, 065 (2015)
  doi:10.1007/JHEP09(2015)065
  [arXiv:1505.04858 [hep-ph]].



\bibitem{King:2013iva} 
  S.~F.~King,
  JHEP {\bf 1307}, 137 (2013)
  doi:10.1007/JHEP07(2013)137
  [arXiv:1304.6264 [hep-ph]].



\bibitem{Bjorkeroth:2014vha} 
  F.~Björkeroth and S.~F.~King,
  J.\ Phys.\ G {\bf 42}, no. 12, 125002 (2015)
  doi:10.1088/0954-3899/42/12/125002
  [arXiv:1412.6996 [hep-ph]].



\bibitem{King:2015dvf} 
  S.~F.~King,
  JHEP {\bf 1602}, 085 (2016)
  doi:10.1007/JHEP02(2016)085
  [arXiv:1512.07531 [hep-ph]].



\bibitem{Bjorkeroth:2015ora} 
  F.~Björkeroth, F.~J.~de Anda, I.~de Medeiros Varzielas and S.~F.~King,
  JHEP {\bf 1506}, 141 (2015)
  doi:10.1007/JHEP06(2015)141
  [arXiv:1503.03306 [hep-ph]].



\bibitem{Bjorkeroth:2015tsa} 
  F.~Björkeroth, F.~J.~de Anda, I.~de Medeiros Varzielas and S.~F.~King,
  JHEP {\bf 1510}, 104 (2015)
  doi:10.1007/JHEP10(2015)104
  [arXiv:1505.05504 [hep-ph]].



\bibitem{King:2016yvg} 
  S.~F.~King and C.~Luhn,
  JHEP {\bf 1609}, 023 (2016)
  doi:10.1007/JHEP09(2016)023
  [arXiv:1607.05276 [hep-ph]].



\bibitem{Ballett:2016yod} 
  P.~Ballett, S.~F.~King, S.~Pascoli, N.~W.~Prouse and T.~Wang,
  JHEP {\bf 1703}, 110 (2017)
  doi:10.1007/JHEP03(2017)110
  [arXiv:1612.01999 [hep-ph]].



\bibitem{King:2018fqh} 
  S.~F.~King, S.~Molina Sedgwick and S.~J.~Rowley,
  JHEP {\bf 1810}, 184 (2018)
  doi:10.1007/JHEP10(2018)184
  [arXiv:1808.01005 [hep-ph]].



\bibitem{King:2018kka} 
  S.~F.~King and C.~C.~Nishi,
  Phys.\ Lett.\ B {\bf 785}, 391 (2018)
  doi:10.1016/j.physletb.2018.08.056
  [arXiv:1807.00023 [hep-ph]].



\bibitem{King:2019tbt} 
  S.~F.~King and Y.~L.~Zhou,
  JHEP {\bf 1905}, 217 (2019)
  doi:10.1007/JHEP05(2019)217
  [arXiv:1901.06877 [hep-ph]].



\bibitem{Abada:2012cq} 
  A.~Abada, D.~Das, A.~Vicente and C.~Weiland,
  JHEP {\bf 1209}, 015 (2012)
  doi:10.1007/JHEP09(2012)015
  [arXiv:1206.6497 [hep-ph]].



\bibitem{Deppisch:2004fa} 
  F.~Deppisch and J.~W.~F.~Valle,
  Phys.\ Rev.\ D {\bf 72}, 036001 (2005)
  doi:10.1103/PhysRevD.72.036001
  [hep-ph/0406040].



\bibitem{Abada:2014kba} 
  A.~Abada, M.~E.~Krauss, W.~Porod, F.~Staub, A.~Vicente and C.~Weiland,
  JHEP {\bf 1411}, 048 (2014)
  doi:10.1007/JHEP11(2014)048
  [arXiv:1408.0138 [hep-ph]].



\bibitem{Abada:2014vea} 
  A.~Abada and M.~Lucente,
  Nucl.\ Phys.\ B {\bf 885}, 651 (2014)
  doi:10.1016/j.nuclphysb.2014.06.003
  [arXiv:1401.1507 [hep-ph]].



\bibitem{Abada:2016awd} 
  A.~Abada and T.~Toma,
  JHEP {\bf 1608}, 079 (2016)
  doi:10.1007/JHEP08(2016)079
  [arXiv:1605.07643 [hep-ph]].



\bibitem{Abada:2018qok} 
  A.~Abada, Á.~Hernández-Cabezudo and X.~Marcano,
  JHEP {\bf 1901}, 041 (2019)
  doi:10.1007/JHEP01(2019)041
  [arXiv:1807.01331 [hep-ph]].



\bibitem{Mohapatra:1986bd} 
  R.~N.~Mohapatra and J.~W.~F.~Valle,
  Phys.\ Rev.\ D {\bf 34}, 1642 (1986).
  doi:10.1103/PhysRevD.34.1642



\bibitem{Bertuzzo:2018ftf} 
  E.~Bertuzzo, S.~Jana, P.~A.~N.~Machado and R.~Zukanovich Funchal,
  Phys.\ Lett.\ B {\bf 791}, 210 (2019)
  doi:10.1016/j.physletb.2019.02.023
  [arXiv:1808.02500 [hep-ph]].



\bibitem{Altarelli:2009gn} 
  G.~Altarelli, F.~Feruglio and L.~Merlo,
  JHEP {\bf 0905}, 020 (2009)
  doi:10.1088/1126-6708/2009/05/020
  [arXiv:0903.1940 [hep-ph]].



\bibitem{Bazzocchi:2009da} 
  F.~Bazzocchi, L.~Merlo and S.~Morisi,
  Phys.\ Rev.\ D {\bf 80}, 053003 (2009)
  doi:10.1103/PhysRevD.80.053003
  [arXiv:0902.2849 [hep-ph]].



\bibitem{Bazzocchi:2009pv} 
  F.~Bazzocchi, L.~Merlo and S.~Morisi,
  Nucl.\ Phys.\ B {\bf 816}, 204 (2009)
  doi:10.1016/j.nuclphysb.2009.03.005
  [arXiv:0901.2086 [hep-ph]].



\bibitem{Toorop:2010yh} 
  R.~de Adelhart Toorop, F.~Bazzocchi and L.~Merlo,
  JHEP {\bf 1008}, 001 (2010)
  doi:10.1007/JHEP08(2010)001
  [arXiv:1003.4502 [hep-ph]].



\bibitem{Patel:2010hr} 
  K.~M.~Patel,
  Phys.\ Lett.\ B {\bf 695}, 225 (2011)
  doi:10.1016/j.physletb.2010.11.024
  [arXiv:1008.5061 [hep-ph]].



\bibitem{Morisi:2011pm} 
  S.~Morisi, K.~M.~Patel and E.~Peinado,
  Phys.\ Rev.\ D {\bf 84}, 053002 (2011)
  doi:10.1103/PhysRevD.84.053002
  [arXiv:1107.0696 [hep-ph]].



\bibitem{Altarelli:2012bn} 
  G.~Altarelli, F.~Feruglio, L.~Merlo and E.~Stamou,
  JHEP {\bf 1208}, 021 (2012)
  doi:10.1007/JHEP08(2012)021
  [arXiv:1205.4670 [hep-ph]].



\bibitem{Mohapatra:2012tb} 
  R.~N.~Mohapatra and C.~C.~Nishi,
  Phys.\ Rev.\ D {\bf 86}, 073007 (2012)
  doi:10.1103/PhysRevD.86.073007
  [arXiv:1208.2875 [hep-ph]].



\bibitem{BhupalDev:2012nm} 
  P.~S.~Bhupal Dev, B.~Dutta, R.~N.~Mohapatra and M.~Severson,
  Phys.\ Rev.\ D {\bf 86}, 035002 (2012)
  doi:10.1103/PhysRevD.86.035002
  [arXiv:1202.4012 [hep-ph]].



\bibitem{Varzielas:2012pa} 
  I.~de Medeiros Varzielas and L.~Lavoura,
  J.\ Phys.\ G {\bf 40}, 085002 (2013)
  doi:10.1088/0954-3899/40/8/085002
  [arXiv:1212.3247 [hep-ph]].



\bibitem{Ding:2013hpa} 
  G.~J.~Ding, S.~F.~King, C.~Luhn and A.~J.~Stuart,
  JHEP {\bf 1305}, 084 (2013)
  doi:10.1007/JHEP05(2013)084
  [arXiv:1303.6180 [hep-ph]].



\bibitem{Ishimori:2010fs} 
  H.~Ishimori, Y.~Shimizu, M.~Tanimoto and A.~Watanabe,
  Phys.\ Rev.\ D {\bf 83}, 033004 (2011)
  doi:10.1103/PhysRevD.83.033004
  [arXiv:1010.3805 [hep-ph]].



\bibitem{Ding:2013eca} 
  G.~J.~Ding and Y.~L.~Zhou,
  Nucl.\ Phys.\ B {\bf 876}, 418 (2013)
  doi:10.1016/j.nuclphysb.2013.08.011
  [arXiv:1304.2645 [hep-ph]].



\bibitem{Hagedorn:2011un} 
  C.~Hagedorn and M.~Serone,
  JHEP {\bf 1110}, 083 (2011)
  doi:10.1007/JHEP10(2011)083
  [arXiv:1106.4021 [hep-ph]].



\bibitem{Campos:2014zaa} 
  M.~D.~Campos, A.~E.~Cárcamo Hernández, H.~Päs and E.~Schumacher,
  Phys.\ Rev.\ D {\bf 91}, no. 11, 116011 (2015)
  doi:10.1103/PhysRevD.91.116011
  [arXiv:1408.1652 [hep-ph]].



\bibitem{Dong:2010zu} 
  P.~V.~Dong, H.~N.~Long, D.~V.~Soa and V.~V.~Vien,
  Eur.\ Phys.\ J.\ C {\bf 71}, 1544 (2011)
  doi:10.1140/epjc/s10052-011-1544-2
  [arXiv:1009.2328 [hep-ph]].



\bibitem{VanVien:2015xha} 
  V.~V.~Vien, H.~N.~Long and D.~P.~Khoi,
  Int.\ J.\ Mod.\ Phys.\ A {\bf 30}, no. 17, 1550102 (2015)
  doi:10.1142/S0217751X1550102X
  [arXiv:1506.06063 [hep-ph]].



\bibitem{deAnda:2017yeb} 
  F.~J.~de Anda, S.~F.~King and E.~Perdomo,
  JHEP {\bf 1712}, 075 (2017)
  Erratum: [JHEP {\bf 1904}, 069 (2019)]
  doi:10.1007/JHEP12(2017)075, 10.1007/JHEP04(2019)069
  [arXiv:1710.03229 [hep-ph]].



\bibitem{deAnda:2018oik} 
  F.~J.~de Anda and S.~F.~King,
  JHEP {\bf 1807}, 057 (2018)
  doi:10.1007/JHEP07(2018)057
  [arXiv:1803.04978 [hep-ph]].



\bibitem{Chen:2019oey} 
  P.~T.~Chen, G.~J.~Ding, S.~F.~King and C.~C.~Li,
  arXiv:1906.11414 [hep-ph].



\bibitem{deMedeirosVarzielas:2019cyj} 
  I.~De Medeiros Varzielas, S.~F.~King and Y.~L.~Zhou,
  arXiv:1906.02208 [hep-ph].



\bibitem{deMedeirosVarzielas:2019hur} 
  I.~De Medeiros Varzielas, M.~Levy and Y.~L.~Zhou,
  Phys.\ Rev.\ D {\bf 100}, no. 3, 035027 (2019)
  doi:10.1103/PhysRevD.100.035027
  [arXiv:1903.10506 [hep-ph]].



\bibitem{CarcamoHernandez:2019kjy} 
  A.~E.~Cárcamo Hernández, M.~González and N.~A.~Neill,
  Phys.\ Rev.\ D {\bf 101}, no. 3, 035005 (2020)
  doi:10.1103/PhysRevD.101.035005
  [arXiv:1906.00978 [hep-ph]].



\bibitem{King:2019vhv} 
  S.~F.~King and Y.~L.~Zhou,
  Phys.\ Rev.\ D {\bf 101}, no. 1, 015001 (2020)
  doi:10.1103/PhysRevD.101.015001
  [arXiv:1908.02770 [hep-ph]].



\bibitem{Ma:2001dn} 
  E.~Ma and G.~Rajasekaran,
  Phys.\ Rev.\ D {\bf 64}, 113012 (2001)
  doi:10.1103/PhysRevD.64.113012
  [hep-ph/0106291].



\bibitem{He:2006dk} 
  X.~G.~He, Y.~Y.~Keum and R.~R.~Volkas,
  JHEP {\bf 0604}, 039 (2006)
  doi:10.1088/1126-6708/2006/04/039
  [hep-ph/0601001].



\bibitem{Feruglio:2008ht} 
  F.~Feruglio, C.~Hagedorn, Y.~Lin and L.~Merlo,
  Nucl.\ Phys.\ B {\bf 809}, 218 (2009)
  doi:10.1016/j.nuclphysb.2008.10.002
  [arXiv:0807.3160 [hep-ph]].



\bibitem{Feruglio:2009hu} 
  F.~Feruglio, C.~Hagedorn, Y.~Lin and L.~Merlo,
  Nucl.\ Phys.\ B {\bf 832}, 251 (2010)
  doi:10.1016/j.nuclphysb.2010.02.010
  [arXiv:0911.3874 [hep-ph]].



\bibitem{Chen:2009um} 
  M.~C.~Chen and S.~F.~King,
  JHEP {\bf 0906}, 072 (2009)
  doi:10.1088/1126-6708/2009/06/072
  [arXiv:0903.0125 [hep-ph]].



\bibitem{Varzielas:2010mp} 
  I.~de Medeiros Varzielas and L.~Merlo,
  JHEP {\bf 1102}, 062 (2011)
  doi:10.1007/JHEP02(2011)062
  [arXiv:1011.6662 [hep-ph]].



\bibitem{Ahn:2012tv} 
  Y.~H.~Ahn and S.~K.~Kang,
  Phys.\ Rev.\ D {\bf 86}, 093003 (2012)
  doi:10.1103/PhysRevD.86.093003
  [arXiv:1203.4185 [hep-ph]].



\bibitem{Memenga:2013vc} 
  N.~Memenga, W.~Rodejohann and H.~Zhang,
  Phys.\ Rev.\ D {\bf 87}, no. 5, 053021 (2013)
  doi:10.1103/PhysRevD.87.053021
  [arXiv:1301.2963 [hep-ph]].



\bibitem{Felipe:2013vwa} 
  R.~Gonzalez Felipe, H.~Serodio and J.~P.~Silva,
  Phys.\ Rev.\ D {\bf 88}, no. 1, 015015 (2013)
  doi:10.1103/PhysRevD.88.015015
  [arXiv:1304.3468 [hep-ph]].



\bibitem{Varzielas:2012ai} 
  I.~de Medeiros Varzielas and D.~Pidt,
  JHEP {\bf 1303}, 065 (2013)
  doi:10.1007/JHEP03(2013)065
  [arXiv:1211.5370 [hep-ph]].



\bibitem{Ishimori:2012fg} 
  H.~Ishimori and E.~Ma,
  Phys.\ Rev.\ D {\bf 86}, 045030 (2012)
  doi:10.1103/PhysRevD.86.045030
  [arXiv:1205.0075 [hep-ph]].



\bibitem{King:2013hj} 
  S.~F.~King, S.~Morisi, E.~Peinado and J.~W.~F.~Valle,
  Phys.\ Lett.\ B {\bf 724}, 68 (2013)
  doi:10.1016/j.physletb.2013.05.067
  [arXiv:1301.7065 [hep-ph]].



\bibitem{Hernandez:2013dta} 
  A.~E.~Carcamo Hernandez, I.~de Medeiros Varzielas, S.~G.~Kovalenko, H.~Päs and I.~Schmidt,
  Phys.\ Rev.\ D {\bf 88}, no. 7, 076014 (2013)
  doi:10.1103/PhysRevD.88.076014
  [arXiv:1307.6499 [hep-ph]].



\bibitem{Babu:2002dz} 
  K.~S.~Babu, E.~Ma and J.~W.~F.~Valle,
  Phys.\ Lett.\ B {\bf 552}, 207 (2003)
  doi:10.1016/S0370-2693(02)03153-2
  [hep-ph/0206292].



\bibitem{Altarelli:2005yx} 
  G.~Altarelli and F.~Feruglio,
  Nucl.\ Phys.\ B {\bf 741}, 215 (2006)
  doi:10.1016/j.nuclphysb.2006.02.015
  [hep-ph/0512103].



\bibitem{Gupta:2011ct} 
  S.~Gupta, A.~S.~Joshipura and K.~M.~Patel,
  Phys.\ Rev.\ D {\bf 85}, 031903 (2012)
  doi:10.1103/PhysRevD.85.031903
  [arXiv:1112.6113 [hep-ph]].



\bibitem{Morisi:2013eca} 
  S.~Morisi, M.~Nebot, K.~M.~Patel, E.~Peinado and J.~W.~F.~Valle,
  Phys.\ Rev.\ D {\bf 88}, 036001 (2013)
  doi:10.1103/PhysRevD.88.036001
  [arXiv:1303.4394 [hep-ph]].



\bibitem{Altarelli:2005yp} 
  G.~Altarelli and F.~Feruglio,
  Nucl.\ Phys.\ B {\bf 720}, 64 (2005)
  doi:10.1016/j.nuclphysb.2005.05.005
  [hep-ph/0504165].



\bibitem{Kadosh:2010rm} 
  A.~Kadosh and E.~Pallante,
  JHEP {\bf 1008}, 115 (2010)
  doi:10.1007/JHEP08(2010)115
  [arXiv:1004.0321 [hep-ph]].



\bibitem{Kadosh:2013nra} 
  A.~Kadosh,
  JHEP {\bf 1306}, 114 (2013)
  doi:10.1007/JHEP06(2013)114
  [arXiv:1303.2645 [hep-ph]].



\bibitem{delAguila:2010vg} 
  F.~del Aguila, A.~Carmona and J.~Santiago,
  JHEP {\bf 1008}, 127 (2010)
  doi:10.1007/JHEP08(2010)127
  [arXiv:1001.5151 [hep-ph]].



\bibitem{Campos:2014lla} 
  M.~D.~Campos, A.~E.~Cárcamo Hernández, S.~Kovalenko, I.~Schmidt and E.~Schumacher,
  Phys.\ Rev.\ D {\bf 90}, no. 1, 016006 (2014)
  doi:10.1103/PhysRevD.90.016006
  [arXiv:1403.2525 [hep-ph]].



\bibitem{Vien:2014pta} 
  V.~V.~Vien and H.~N.~Long,
  Int.\ J.\ Mod.\ Phys.\ A {\bf 30}, no. 21, 1550117 (2015)
  doi:10.1142/S0217751X15501171
  [arXiv:1405.4665 [hep-ph]].



\bibitem{Joshipura:2015dsa} 
  A.~S.~Joshipura and K.~M.~Patel,
  Phys.\ Lett.\ B {\bf 749}, 159 (2015)
  doi:10.1016/j.physletb.2015.07.062
  [arXiv:1507.01235 [hep-ph]].



\bibitem{Hernandez:2015tna} 
  A.~E.~Cárcamo Hernández and R.~Martinez,
  Nucl.\ Phys.\ B {\bf 905}, 337 (2016)
  doi:10.1016/j.nuclphysb.2016.02.025
  [arXiv:1501.05937 [hep-ph]].



\bibitem{Karmakar:2016cvb} 
  B.~Karmakar and A.~Sil,
  Phys.\ Rev.\ D {\bf 96}, no. 1, 015007 (2017)
  doi:10.1103/PhysRevD.96.015007
  [arXiv:1610.01909 [hep-ph]].



\bibitem{Chattopadhyay:2017zvs} 
  P.~Chattopadhyay and K.~M.~Patel,
  Nucl.\ Phys.\ B {\bf 921}, 487 (2017)
  doi:10.1016/j.nuclphysb.2017.06.008
  [arXiv:1703.09541 [hep-ph]].



\bibitem{CarcamoHernandez:2017kra} 
  A.~E.~Cárcamo Hernández and H.~N.~Long,
  J.\ Phys.\ G {\bf 45}, no. 4, 045001 (2018)
  doi:10.1088/1361-6471/aaace7
  [arXiv:1705.05246 [hep-ph]].



\bibitem{Ma:2017moj} 
  E.~Ma and G.~Rajasekaran,
  EPL {\bf 119}, no. 3, 31001 (2017)
  doi:10.1209/0295-5075/119/31001
  [arXiv:1708.02208 [hep-ph]].



\bibitem{CentellesChulia:2017koy} 
  S.~Centelles Chuliá, R.~Srivastava and J.~W.~F.~Valle,
  Phys.\ Lett.\ B {\bf 773}, 26 (2017)
  doi:10.1016/j.physletb.2017.07.065
  [arXiv:1706.00210 [hep-ph]].



\bibitem{Bjorkeroth:2017tsz} 
  F.~Björkeroth, E.~J.~Chun and S.~F.~King,
  Phys.\ Lett.\ B {\bf 777}, 428 (2018)
  doi:10.1016/j.physletb.2017.12.058
  [arXiv:1711.05741 [hep-ph]].



\bibitem{Srivastava:2017sno} 
  R.~Srivastava, C.~A.~Ternes, M.~Tórtola and J.~W.~F.~Valle,
  Phys.\ Lett.\ B {\bf 778}, 459 (2018)
  doi:10.1016/j.physletb.2018.01.014
  [arXiv:1711.10318 [hep-ph]].



\bibitem{Borah:2017dmk} 
  D.~Borah and B.~Karmakar,
  Phys.\ Lett.\ B {\bf 780}, 461 (2018)
  doi:10.1016/j.physletb.2018.03.047
  [arXiv:1712.06407 [hep-ph]].



\bibitem{Belyaev:2018vkl} 
  A.~S.~Belyaev, S.~F.~King and P.~B.~Schaefers,
  Phys.\ Rev.\ D {\bf 97}, no. 11, 115002 (2018)
  doi:10.1103/PhysRevD.97.115002
  [arXiv:1801.00514 [hep-ph]].



\bibitem{CarcamoHernandez:2018aon} 
  A.~E.~Cárcamo Hernández and S.~F.~King,
  Phys.\ Rev.\ D {\bf 99}, no. 9, 095003 (2019)
  doi:10.1103/PhysRevD.99.095003
  [arXiv:1803.07367 [hep-ph]].



\bibitem{Srivastava:2018ser} 
  R.~Srivastava, C.~A.~Ternes, M.~Tórtola and J.~W.~F.~Valle,
  Phys.\ Rev.\ D {\bf 97}, no. 9, 095025 (2018)
  doi:10.1103/PhysRevD.97.095025
  [arXiv:1803.10247 [hep-ph]].



\bibitem{delaVega:2018cnx} 
  L.~M.~G.~De La Vega, R.~Ferro-Hernandez and E.~Peinado,
  Phys.\ Rev.\ D {\bf 99}, no. 5, 055044 (2019)
  doi:10.1103/PhysRevD.99.055044
  [arXiv:1811.10619 [hep-ph]].



\bibitem{Borah:2018nvu} 
  D.~Borah and B.~Karmakar,
  Phys.\ Lett.\ B {\bf 789}, 59 (2019)
  doi:10.1016/j.physletb.2018.12.006
  [arXiv:1806.10685 [hep-ph]].



\bibitem{Pramanick:2019qpg} 
  S.~Pramanick,
  arXiv:1903.04208 [hep-ph].



\bibitem{CarcamoHernandez:2019pmy} 
  A.~E.~Cárcamo Hernández, J.~Marchant González and U.~J.~Saldaña-Salazar,
  Phys.\ Rev.\ D {\bf 100}, no. 3, 035024 (2019)
  doi:10.1103/PhysRevD.100.035024
  [arXiv:1904.09993 [hep-ph]].



\bibitem{Luhn:2007sy} 
  C.~Luhn, S.~Nasri and P.~Ramond,
  Phys.\ Lett.\ B {\bf 652}, 27 (2007)
  doi:10.1016/j.physletb.2007.06.059
  [arXiv:0706.2341 [hep-ph]].



\bibitem{Hagedorn:2008bc} 
  C.~Hagedorn, M.~A.~Schmidt and A.~Y.~Smirnov,
  Phys.\ Rev.\ D {\bf 79}, 036002 (2009)
  doi:10.1103/PhysRevD.79.036002
  [arXiv:0811.2955 [hep-ph]].



\bibitem{Cao:2010mp} 
  Q.~H.~Cao, S.~Khalil, E.~Ma and H.~Okada,
  Phys.\ Rev.\ Lett.\  {\bf 106}, 131801 (2011)
  doi:10.1103/PhysRevLett.106.131801
  [arXiv:1009.5415 [hep-ph]].



\bibitem{Luhn:2012bc} 
  C.~Luhn, K.~M.~Parattu and A.~Wingerter,
  JHEP {\bf 1212}, 096 (2012)
  doi:10.1007/JHEP12(2012)096
  [arXiv:1210.1197 [hep-ph]].



\bibitem{Kajiyama:2013lja} 
  Y.~Kajiyama, H.~Okada and K.~Yagyu,
  JHEP {\bf 1310}, 196 (2013)
  doi:10.1007/JHEP10(2013)196
  [arXiv:1307.0480 [hep-ph]].



\bibitem{Bonilla:2014xla} 
  C.~Bonilla, S.~Morisi, E.~Peinado and J.~W.~F.~Valle,
  Phys.\ Lett.\ B {\bf 742}, 99 (2015)
  doi:10.1016/j.physletb.2015.01.017
  [arXiv:1411.4883 [hep-ph]].



\bibitem{Vien:2014gza} 
  V.~V.~Vien and H.~N.~Long,
  JHEP {\bf 1404}, 133 (2014)
  doi:10.1007/JHEP04(2014)133
  [arXiv:1402.1256 [hep-ph]].



\bibitem{Vien:2015koa} 
  V.~V.~Vien,
  Mod.\ Phys.\ Lett.\ A {\bf 29}, 28 (2014)
  doi:10.1142/S0217732314501399
  [arXiv:1508.02585 [hep-ph]].



\bibitem{Hernandez:2015cra} 
  A.~E.~Cárcamo Hernández and R.~Martinez,
  J.\ Phys.\ G {\bf 43}, no. 4, 045003 (2016)
  doi:10.1088/0954-3899/43/4/045003
  [arXiv:1501.07261 [hep-ph]].



\bibitem{Arbelaez:2015toa} 
  C.~Arbeláez, A.~E.~Cárcamo Hernández, S.~Kovalenko and I.~Schmidt,
  Phys.\ Rev.\ D {\bf 92}, no. 11, 115015 (2015)
  doi:10.1103/PhysRevD.92.115015
  [arXiv:1507.03852 [hep-ph]].



\bibitem{Branco:1983tn} 
  G.~C.~Branco, J.~M.~Gerard and W.~Grimus,
  Phys.\ Lett.\  {\bf 136B}, 383 (1984).
  doi:10.1016/0370-2693(84)92024-0



\bibitem{deMedeirosVarzielas:2006fc} 
  I.~de Medeiros Varzielas, S.~F.~King and G.~G.~Ross,
  Phys.\ Lett.\ B {\bf 648}, 201 (2007)
  doi:10.1016/j.physletb.2007.03.009
  [hep-ph/0607045].



\bibitem{Ma:2007wu} 
  E.~Ma,
  Phys.\ Lett.\ B {\bf 660}, 505 (2008)
  doi:10.1016/j.physletb.2007.12.060
  [arXiv:0709.0507 [hep-ph]].



\bibitem{Varzielas:2012nn} 
  I.~de Medeiros Varzielas, D.~Emmanuel-Costa and P.~Leser,
  Phys.\ Lett.\ B {\bf 716}, 193 (2012)
  doi:10.1016/j.physletb.2012.08.008
  [arXiv:1204.3633 [hep-ph]].



\bibitem{Bhattacharyya:2012pi} 
  G.~Bhattacharyya, I.~de Medeiros Varzielas and P.~Leser,
  Phys.\ Rev.\ Lett.\  {\bf 109}, 241603 (2012)
  doi:10.1103/PhysRevLett.109.241603
  [arXiv:1210.0545 [hep-ph]].



\bibitem{Ferreira:2012ri} 
  P.~M.~Ferreira, W.~Grimus, L.~Lavoura and P.~O.~Ludl,
  JHEP {\bf 1209}, 128 (2012)
  doi:10.1007/JHEP09(2012)128
  [arXiv:1206.7072 [hep-ph]].



\bibitem{Ma:2013xqa} 
  E.~Ma,
  Phys.\ Lett.\ B {\bf 723}, 161 (2013)
  doi:10.1016/j.physletb.2013.05.011
  [arXiv:1304.1603 [hep-ph]].



\bibitem{Nishi:2013jqa} 
  C.~C.~Nishi,
  Phys.\ Rev.\ D {\bf 88}, no. 3, 033010 (2013)
  doi:10.1103/PhysRevD.88.033010
  [arXiv:1306.0877 [hep-ph]].



\bibitem{Varzielas:2013sla} 
  I.~de Medeiros Varzielas and D.~Pidt,
  J.\ Phys.\ G {\bf 41}, 025004 (2014)
  doi:10.1088/0954-3899/41/2/025004
  [arXiv:1307.0711 [hep-ph]].



\bibitem{Aranda:2013gga} 
  A.~Aranda, C.~Bonilla, S.~Morisi, E.~Peinado and J.~W.~F.~Valle,
  Phys.\ Rev.\ D {\bf 89}, no. 3, 033001 (2014)
  doi:10.1103/PhysRevD.89.033001
  [arXiv:1307.3553 [hep-ph]].



\bibitem{Harrison:2014jqa} 
  P.~F.~Harrison, R.~Krishnan and W.~G.~Scott,
  Int.\ J.\ Mod.\ Phys.\ A {\bf 29}, no. 18, 1450095 (2014)
  doi:10.1142/S0217751X1450095X
  [arXiv:1406.2025 [hep-ph]].



\bibitem{Ma:2014eka} 
  E.~Ma and A.~Natale,
  Phys.\ Lett.\ B {\bf 734}, 403 (2014)
  doi:10.1016/j.physletb.2014.05.070
  [arXiv:1403.6772 [hep-ph]].



\bibitem{Abbas:2014ewa} 
  M.~Abbas and S.~Khalil,
  Phys.\ Rev.\ D {\bf 91}, no. 5, 053003 (2015)
  doi:10.1103/PhysRevD.91.053003
  [arXiv:1406.6716 [hep-ph]].



\bibitem{Abbas:2015zna} 
  M.~Abbas, S.~Khalil, A.~Rashed and A.~Sil,
  Phys.\ Rev.\ D {\bf 93}, no. 1, 013018 (2016)
  doi:10.1103/PhysRevD.93.013018
  [arXiv:1508.03727 [hep-ph]].



\bibitem{Varzielas:2015aua} 
  I.~de Medeiros Varzielas,
  JHEP {\bf 1508}, 157 (2015)
  doi:10.1007/JHEP08(2015)157
  [arXiv:1507.00338 [hep-ph]].



\bibitem{Bjorkeroth:2015uou} 
  F.~Björkeroth, F.~J.~de Anda, I.~de Medeiros Varzielas and S.~F.~King,
  Phys.\ Rev.\ D {\bf 94}, no. 1, 016006 (2016)
  doi:10.1103/PhysRevD.94.016006
  [arXiv:1512.00850 [hep-ph]].



\bibitem{Chen:2015jta} 
  P.~Chen, G.~J.~Ding, A.~D.~Rojas, C.~A.~Vaquera-Araujo and J.~W.~F.~Valle,
  JHEP {\bf 1601}, 007 (2016)
  doi:10.1007/JHEP01(2016)007
  [arXiv:1509.06683 [hep-ph]].



\bibitem{Vien:2016tmh} 
  V.~V.~Vien, A.~E.~Cárcamo Hernández and H.~N.~Long,
  Nucl.\ Phys.\ B {\bf 913}, 792 (2016)
  doi:10.1016/j.nuclphysb.2016.10.010
  [arXiv:1601.03300 [hep-ph]].



\bibitem{Hernandez:2016eod} 
  A.~E.~Cárcamo Hernández, H.~N.~Long and V.~V.~Vien,
  Eur.\ Phys.\ J.\ C {\bf 76}, no. 5, 242 (2016)
  doi:10.1140/epjc/s10052-016-4074-0
  [arXiv:1601.05062 [hep-ph]].



\bibitem{CarcamoHernandez:2017owh} 
  A.~E.~Cárcamo Hernández, S.~Kovalenko, J.~W.~F.~Valle and C.~A.~Vaquera-Araujo,
  JHEP {\bf 1707}, 118 (2017)
  doi:10.1007/JHEP07(2017)118
  [arXiv:1705.06320 [hep-ph]].



\bibitem{deMedeirosVarzielas:2017sdv} 
  I.~de Medeiros Varzielas, G.~G.~Ross and J.~Talbert,
  JHEP {\bf 1803}, 007 (2018)
  doi:10.1007/JHEP03(2018)007
  [arXiv:1710.01741 [hep-ph]].



\bibitem{Bernal:2017xat} 
  N.~Bernal, A.~E.~Cárcamo Hernández, I.~de Medeiros Varzielas and S.~Kovalenko,
  JHEP {\bf 1805}, 053 (2018)
  doi:10.1007/JHEP05(2018)053
  [arXiv:1712.02792 [hep-ph]].



\bibitem{CarcamoHernandez:2018iel} 
  A.~E.~Cárcamo Hernández, H.~N.~Long and V.~V.~Vien,
  Eur.\ Phys.\ J.\ C {\bf 78}, no. 10, 804 (2018)
  doi:10.1140/epjc/s10052-018-6284-0
  [arXiv:1803.01636 [hep-ph]].



\bibitem{deMedeirosVarzielas:2018vab} 
  I.~De Medeiros Varzielas, M.~L.~López-Ibáñez, A.~Melis and O.~Vives,
  JHEP {\bf 1809}, 047 (2018)
  doi:10.1007/JHEP09(2018)047
  [arXiv:1807.00860 [hep-ph]].



\bibitem{CarcamoHernandez:2018hst} 
  A.~E.~Cárcamo Hernández, S.~Kovalenko, J.~W.~F.~Valle and C.~A.~Vaquera-Araujo,
  JHEP {\bf 1902}, 065 (2019)
  doi:10.1007/JHEP02(2019)065
  [arXiv:1811.03018 [hep-ph]].



\bibitem{Aranda:2000tm} 
  A.~Aranda, C.~D.~Carone and R.~F.~Lebed,
  Phys.\ Rev.\ D {\bf 62}, 016009 (2000)
  doi:10.1103/PhysRevD.62.016009
  [hep-ph/0002044].



\bibitem{Feruglio:2007uu} 
  F.~Feruglio, C.~Hagedorn, Y.~Lin and L.~Merlo,
  Nucl.\ Phys.\ B {\bf 775}, 120 (2007)
  Erratum: [Nucl.\ Phys.\ B {\bf 836}, 127 (2010)]
  doi:10.1016/j.nuclphysb.2007.04.002, 10.1016/j.nuclphysb.2010.04.018
  [hep-ph/0702194].



\bibitem{Sen:2007vx} 
  S.~Sen,
  Phys.\ Rev.\ D {\bf 76}, 115020 (2007)
  doi:10.1103/PhysRevD.76.115020
  [arXiv:0710.2734 [hep-ph]].



\bibitem{Aranda:2007dp} 
  A.~Aranda,
  Phys.\ Rev.\ D {\bf 76}, 111301 (2007)
  doi:10.1103/PhysRevD.76.111301
  [arXiv:0707.3661 [hep-ph]].



\bibitem{Chen:2007afa} 
  M.~C.~Chen and K.~T.~Mahanthappa,
  Phys.\ Lett.\ B {\bf 652}, 34 (2007)
  doi:10.1016/j.physletb.2007.06.064
  [arXiv:0705.0714 [hep-ph]].



\bibitem{Eby:2008uc} 
  D.~A.~Eby, P.~H.~Frampton and S.~Matsuzaki,
  Phys.\ Lett.\ B {\bf 671}, 386 (2009)
  doi:10.1016/j.physletb.2008.11.074
  [arXiv:0810.4899 [hep-ph]].



\bibitem{Frampton:2008bz} 
  P.~H.~Frampton, T.~W.~Kephart and S.~Matsuzaki,
  Phys.\ Rev.\ D {\bf 78}, 073004 (2008)
  doi:10.1103/PhysRevD.78.073004
  [arXiv:0807.4713 [hep-ph]].



\bibitem{Frampton:2008ep} 
  P.~H.~Frampton and S.~Matsuzaki,
  Mod.\ Phys.\ Lett.\ A {\bf 24}, 429 (2009)
  doi:10.1142/S0217732309030229
  [arXiv:0807.4785 [hep-ph]].



\bibitem{Eby:2009ii} 
  D.~A.~Eby, P.~H.~Frampton and S.~Matsuzaki,
  Phys.\ Rev.\ D {\bf 80}, 053007 (2009)
  doi:10.1103/PhysRevD.80.053007
  [arXiv:0907.3425 [hep-ph]].



\bibitem{Frampton:2009fw} 
  P.~H.~Frampton and S.~Matsuzaki,
  Phys.\ Lett.\ B {\bf 679}, 347 (2009)
  doi:10.1016/j.physletb.2009.08.001
  [arXiv:0902.1140 [hep-ph]].



\bibitem{Eby:2011ph} 
  D.~A.~Eby, P.~H.~Frampton, X.~G.~He and T.~W.~Kephart,
  Phys.\ Rev.\ D {\bf 84}, 037302 (2011)
  doi:10.1103/PhysRevD.84.037302
  [arXiv:1103.5737 [hep-ph]].



\bibitem{Eby:2011qa} 
  D.~A.~Eby and P.~H.~Frampton,
  Phys.\ Lett.\ B {\bf 713}, 249 (2012)
  doi:10.1016/j.physletb.2012.06.004
  [arXiv:1111.4938 [hep-ph]].



\bibitem{Chen:2011tj} 
  M.~C.~Chen and K.~T.~Mahanthappa,
  arXiv:1107.3856 [hep-ph].



\bibitem{Meroni:2012ty} 
  A.~Meroni, S.~T.~Petcov and M.~Spinrath,
  Phys.\ Rev.\ D {\bf 86}, 113003 (2012)
  doi:10.1103/PhysRevD.86.113003
  [arXiv:1205.5241 [hep-ph]].



\bibitem{Frampton:2013lva} 
  P.~H.~Frampton, C.~M.~Ho and T.~W.~Kephart,
  Phys.\ Rev.\ D {\bf 89}, no. 2, 027701 (2014)
  doi:10.1103/PhysRevD.89.027701
  [arXiv:1305.4402 [hep-ph]].



\bibitem{Chen:2013wba} 
  M.~C.~Chen, J.~Huang, K.~T.~Mahanthappa and A.~M.~Wijangco,
  JHEP {\bf 1310}, 112 (2013)
  doi:10.1007/JHEP10(2013)112
  [arXiv:1307.7711 [hep-ph]].



\bibitem{Girardi:2013sza} 
  I.~Girardi, A.~Meroni, S.~T.~Petcov and M.~Spinrath,
  JHEP {\bf 1402}, 050 (2014)
  doi:10.1007/JHEP02(2014)050
  [arXiv:1312.1966 [hep-ph]].



\bibitem{Carone:2016xsi} 
  C.~D.~Carone, S.~Chaurasia and S.~Vasquez,
  Phys.\ Rev.\ D {\bf 95}, no. 1, 015025 (2017)
  doi:10.1103/PhysRevD.95.015025
  [arXiv:1611.00784 [hep-ph]].



\bibitem{Vien:2018otl} 
  V.~V.~Vien, H.~N.~Long and A.~E.~Cárcamo Hernández,
  Mod.\ Phys.\ Lett.\ A {\bf 34}, no. 01, 1950005 (2019)
  doi:10.1142/S0217732319500056
  [arXiv:1812.07263 [hep-ph]].



\bibitem{Carone:2019lfc} 
  C.~D.~Carone and M.~Merchand,
  Phys.\ Rev.\ D {\bf 100}, no. 3, 035006 (2019)
  doi:10.1103/PhysRevD.100.035006
  [arXiv:1904.11059 [hep-ph]].



\bibitem{CarcamoHernandez:2019vih} 
  A.~E.~Cárcamo Hernández, Y.~Hidalgo Velásquez and N.~A.~Pérez-Julve,
  Eur.\ Phys.\ J.\ C {\bf 79}, no. 10, 828 (2019)
  doi:10.1140/epjc/s10052-019-7325-z
  [arXiv:1905.02323 [hep-ph]].



\bibitem{Ding:2019zxk} 
  G.~J.~Ding, S.~F.~King and X.~G.~Liu,
  JHEP {\bf 1909}, 074 (2019)
  doi:10.1007/JHEP09(2019)074
  [arXiv:1907.11714 [hep-ph]].



\bibitem{GonzalezGarcia:1988rw} 
  M.~C.~Gonzalez-Garcia and J.~W.~F.~Valle,
  Phys.\ Lett.\ B {\bf 216}, 360 (1989).
  doi:10.1016/0370-2693(89)91131-3



\bibitem{Akhmedov:1995vm} 
  E.~K.~Akhmedov, M.~Lindner, E.~Schnapka and J.~W.~F.~Valle,
  Phys.\ Rev.\ D {\bf 53}, 2752 (1996)
  doi:10.1103/PhysRevD.53.2752
  [hep-ph/9509255].



\bibitem{Akhmedov:1995ip} 
  E.~K.~Akhmedov, M.~Lindner, E.~Schnapka and J.~W.~F.~Valle,
  Phys.\ Lett.\ B {\bf 368}, 270 (1996)
  doi:10.1016/0370-2693(95)01504-3
  [hep-ph/9507275].



\bibitem{Malinsky:2005bi} 
  M.~Malinsky, J.~C.~Romao and J.~W.~F.~Valle,
  Phys.\ Rev.\ Lett.\  {\bf 95}, 161801 (2005)
  doi:10.1103/PhysRevLett.95.161801
  [hep-ph/0506296].



\bibitem{Malinsky:2009df} 
  M.~Malinsky, T.~Ohlsson, Z.~z.~Xing and H.~Zhang,
  Phys.\ Lett.\ B {\bf 679}, 242 (2009)
  doi:10.1016/j.physletb.2009.07.038
  [arXiv:0905.2889 [hep-ph]].



\bibitem{Catano:2012kw} 
  M.~E.~Catano, R.~Martinez and F.~Ochoa,
  Phys.\ Rev.\ D {\bf 86}, 073015 (2012)
  doi:10.1103/PhysRevD.86.073015
  [arXiv:1206.1966 [hep-ph]].



\bibitem{deSalas:2017kay} 
  P.~F.~de Salas, D.~V.~Forero, C.~A.~Ternes, M.~Tortola and J.~W.~F.~Valle,
  Phys.\ Lett.\ B {\bf 782}, 633 (2018)
  doi:10.1016/j.physletb.2018.06.019
  [arXiv:1708.01186 [hep-ph]].



\bibitem{Esteban:2018azc} 
  I.~Esteban, M.~C.~Gonzalez-Garcia, A.~Hernandez-Cabezudo, M.~Maltoni and T.~Schwetz,
  JHEP {\bf 1901}, 106 (2019)
  doi:10.1007/JHEP01(2019)106
  [arXiv:1811.05487 [hep-ph]].



\bibitem{Ilakovac:1994kj} 
  A.~Ilakovac and A.~Pilaftsis,
  Nucl.\ Phys.\ B {\bf 437}, 491 (1995)
  doi:10.1016/0550-3213(94)00567-X
  [hep-ph/9403398].



\bibitem{Lindner:2016bgg} 
  M.~Lindner, M.~Platscher and F.~S.~Queiroz,
  Phys.\ Rept.\  {\bf 731}, 1 (2018)
  doi:10.1016/j.physrep.2017.12.001
  [arXiv:1610.06587 [hep-ph]].



\bibitem{Carena:2015uoe} 
  M.~Carena, J.~Ellis, J.~S.~Lee, A.~Pilaftsis and C.~E.~M.~Wagner,
  JHEP {\bf 1602}, 123 (2016)
  doi:10.1007/JHEP02(2016)123
  [arXiv:1512.00437 [hep-ph]].



\bibitem{Kuno:1999jp} 
  Y.~Kuno and Y.~Okada,
  Rev.\ Mod.\ Phys.\  {\bf 73}, 151 (2001)
  doi:10.1103/RevModPhys.73.151
  [hep-ph/9909265].



\bibitem{Aysto:2001zs} 
  J.~Aysto {\it et al.},
  CERN Yellow Report CERN-2004-002, pp.259-306
  doi:10.5170/CERN-2004-002.259
  [hep-ph/0109217].



\bibitem{Bernstein:2013hba} 
  R.~H.~Bernstein and P.~S.~Cooper,
  Phys.\ Rept.\  {\bf 532}, 27 (2013)
  doi:10.1016/j.physrep.2013.07.002
  [arXiv:1307.5787 [hep-ex]].



\bibitem{Ishimori:2010au} 
  H.~Ishimori, T.~Kobayashi, H.~Ohki, Y.~Shimizu, H.~Okada and M.~Tanimoto,
  Prog.\ Theor.\ Phys.\ Suppl.\  {\bf 183}, 1 (2010)
  doi:10.1143/PTPS.183.1
  [arXiv:1003.3552 [hep-th]].
\end{thebibliography}
\end{document}